\pdfoutput=1

\documentclass[12pt,a4paper]{article}

\usepackage{ifthen} 
\newboolean{pdflatex}
\setboolean{pdflatex}{true} 

\newboolean{articletitles}
\setboolean{articletitles}{true} 

\newboolean{uprightparticles}
\setboolean{uprightparticles}{false} 

\newboolean{inbibliography}
\setboolean{inbibliography}{false} 

\usepackage[top=1in, bottom=1.25in, left=1in, right=1in]{geometry}

\usepackage{microtype}
\usepackage{lineno}  
\usepackage{xspace} 
\usepackage{caption} 

\usepackage{graphicx}  
\usepackage{color}
\usepackage{colortbl}
\graphicspath{{./figs/}} 

\usepackage{amsmath} 
\usepackage{amssymb}
\usepackage{amsfonts}
\usepackage{upgreek} 

\newcommand*\patchAmsMathEnvironmentForLineno[1]{%
\expandafter\let\csname old#1\expandafter\endcsname\csname #1\endcsname
\expandafter\let\csname oldend#1\expandafter\endcsname\csname
end#1\endcsname
 \renewenvironment{#1}%
   {\linenomath\csname old#1\endcsname}%
   {\csname oldend#1\endcsname\endlinenomath}%
}
\newcommand*\patchBothAmsMathEnvironmentsForLineno[1]{%
  \patchAmsMathEnvironmentForLineno{#1}%
  \patchAmsMathEnvironmentForLineno{#1*}%
}
\AtBeginDocument{%
\patchBothAmsMathEnvironmentsForLineno{equation}%
\patchBothAmsMathEnvironmentsForLineno{align}%
\patchBothAmsMathEnvironmentsForLineno{flalign}%
\patchBothAmsMathEnvironmentsForLineno{alignat}%
\patchBothAmsMathEnvironmentsForLineno{gather}%
\patchBothAmsMathEnvironmentsForLineno{multline}%
\patchBothAmsMathEnvironmentsForLineno{eqnarray}%
}

\usepackage{hyperref}    
\usepackage[all]{hypcap} 

\usepackage{xspace} 
\usepackage{upgreek}

\def\lhcb {\mbox{LHCb}\xspace}

\def\MagUp {\mbox{\em Mag\kern -0.05em Up}\xspace}

\ifthenelse{\boolean{uprightparticles}}%
{

 \def\Peta        {\ensuremath{\upeta}\xspace}

 \def\Ppi         {\ensuremath{\uppi}\xspace}

 \def\Pphi        {\ensuremath{\upphi}\xspace}

 \def\Ppsi        {\ensuremath{\uppsi}\xspace}                 
 \def\Pomega      {\ensuremath{\upomega}\xspace}                 

 \def\PDelta      {\ensuremath{\Delta}\xspace}                 
 \def\PXi      {\ensuremath{\Xi}\xspace}                 
 \def\PLambda      {\ensuremath{\Lambda}\xspace}                 
 \def\PSigma      {\ensuremath{\Sigma}\xspace}                 
 \def\POmega      {\ensuremath{\Omega}\xspace}                 
 \def\PUpsilon      {\ensuremath{\Upsilon}\xspace}                 
 
 \def\PB      {\ensuremath{\mathrm{B}}\xspace}                 
                  
 \def\PD      {\ensuremath{\mathrm{D}}\xspace}

 \def\PJ      {\ensuremath{\mathrm{J}}\xspace}                 
 \def\PK      {\ensuremath{\mathrm{K}}\xspace}

 \def\Pa      {\ensuremath{\mathrm{a}}\xspace}                 
 \def\Pb      {\ensuremath{\mathrm{b}}\xspace}                 
 \def\Pc      {\ensuremath{\mathrm{c}}\xspace}

 \def\Pf      {\ensuremath{\mathrm{f}}\xspace}

 \def\Pi      {\ensuremath{\mathrm{i}}\xspace}

 \def\Ps      {\ensuremath{\mathrm{s}}\xspace}

}
{

 \def\Peta        {\ensuremath{\eta}\xspace}

 \def\Ppi         {\ensuremath{\pi}\xspace}

 \def\Pphi        {\ensuremath{\phi}\xspace}

 \def\Ppsi        {\ensuremath{\psi}\xspace}                 
 \def\Pomega      {\ensuremath{\omega}\xspace}                 
 \mathchardef\PDelta="7101
 \mathchardef\PXi="7104
 \mathchardef\PLambda="7103
 \mathchardef\PSigma="7106
 \mathchardef\POmega="710A
 \mathchardef\PUpsilon="7107
                  
 \def\PB      {\ensuremath{B}\xspace}                 
                  
 \def\PD      {\ensuremath{D}\xspace}

 \def\PJ      {\ensuremath{J}\xspace}                 
 \def\PK      {\ensuremath{K}\xspace}

 \def\Pa      {\ensuremath{a}\xspace}                 
 \def\Pb      {\ensuremath{b}\xspace}                 
 \def\Pc      {\ensuremath{c}\xspace}

 \def\Pf      {\ensuremath{f}\xspace}

 \def\Pi      {\ensuremath{i}\xspace}

 \def\Ps      {\ensuremath{s}\xspace}

}

\makeatletter
\ifcase \@ptsize \relax
  \newcommand{\miniscule}{\@setfontsize\miniscule{4}{5}}
\or
  \newcommand{\miniscule}{\@setfontsize\miniscule{5}{6}}
\or
  \newcommand{\miniscule}{\@setfontsize\miniscule{5}{6}}
\fi
\makeatother

\DeclareRobustCommand{\optbar}[1]{\shortstack{{\miniscule (\rule[.5ex]{1.25em}{.18mm})}
  \\ [-.7ex] $#1$}}

\def\squark    {{\ensuremath{\Ps}}\xspace}

\def\cquark    {{\ensuremath{\Pc}}\xspace}

\def\bquark    {{\ensuremath{\Pb}}\xspace}

\def\pion   {{\ensuremath{\Ppi}}\xspace}

\def\pip    {{\ensuremath{\pion^+}}\xspace}
\def\pim    {{\ensuremath{\pion^-}}\xspace}

\def\kaon    {{\ensuremath{\PK}}\xspace}
  \def\Kbar    {{\kern 0.2em\overline{\kern -0.2em \PK}{}}\xspace}

\def\KorKbar    {\kern 0.18em\optbar{\kern -0.18em K}{}\xspace}
\def\Kz      {{\ensuremath{\kaon^0}}\xspace}

\def\Kp      {{\ensuremath{\kaon^+}}\xspace}
\def\Km      {{\ensuremath{\kaon^-}}\xspace}

\def\Kstar   {{\ensuremath{\kaon^*}}\xspace}

  \def\Dbar    {{\kern 0.2em\overline{\kern -0.2em \PD}{}}\xspace}
\def\D       {{\ensuremath{\PD}}\xspace}

\def\DorDbar    {\kern 0.18em\optbar{\kern -0.18em D}{}\xspace}

\def\Ds      {{\ensuremath{\D^+_\squark}}\xspace}
\def\Dsp     {{\ensuremath{\D^+_\squark}}\xspace}
\def\Dsm     {{\ensuremath{\D^-_\squark}}\xspace}

\def\B       {{\ensuremath{\PB}}\xspace}
\def\Bbar    {{\ensuremath{\kern 0.18em\overline{\kern -0.18em \PB}{}}}\xspace}

\def\BorBbar    {\kern 0.18em\optbar{\kern -0.18em B}{}\xspace}

\def\Bd      {{\ensuremath{\B^0}}\xspace}
\def\Bs      {{\ensuremath{\B^0_\squark}}\xspace}
\def\Bsb     {{\ensuremath{\Bbar{}^0_\squark}}\xspace}

\def\jpsi     {{\ensuremath{{\PJ\mskip -3mu/\mskip -2mu\Ppsi\mskip 2mu}}}\xspace}

\def\etac     {{\ensuremath{\Peta_\cquark}}\xspace}

  \def\Y#1S{\ensuremath{\PUpsilon{(#1S)}}\xspace}

\def\Lbar        {{\ensuremath{\kern 0.1em\overline{\kern -0.1em\PLambda}}}\xspace}
\def\LorLbar    {\kern 0.18em\optbar{\kern -0.18em \PLambda}{}\xspace}

\def\BF         {{\ensuremath{\mathcal{B}}}\xspace}

\def\BR         {\BF}
\newcommand{\decay}[2]{\ensuremath{#1\!\to #2}\xspace}         

\def\to                 {\ensuremath{\rightarrow}\xspace}

\def\CP                {{\ensuremath{C\!P}}\xspace}

\newcommand{\phis}{{\ensuremath{\phi_{\squark}}}\xspace}

\def\BsToJPsiPhi  {\decay{\Bs}{\jpsi\phi}}

\def\AT#1     {\ensuremath{A_{\mathrm{T}}^{#1}}\xspace}           

\def\C#1      {\ensuremath{\mathcal{C}_{#1}}\xspace}                       
\def\Cp#1     {\ensuremath{\mathcal{C}_{#1}^{'}}\xspace}                    
\def\Ceff#1   {\ensuremath{\mathcal{C}_{#1}^{\mathrm{(eff)}}}\xspace}        
\def\Cpeff#1  {\ensuremath{\mathcal{C}_{#1}^{'\mathrm{(eff)}}}\xspace}       
\def\Ope#1    {\ensuremath{\mathcal{O}_{#1}}\xspace}                       
\def\Opep#1   {\ensuremath{\mathcal{O}_{#1}^{'}}\xspace}                    

\newcommand{\unit}[1]{\ensuremath{\mathrm{ \,#1}}\xspace}          

\newcommand{\tev}{\ifthenelse{\boolean{inbibliography}}{\ensuremath{~T\kern -0.05em eV}\xspace}{\ensuremath{\mathrm{\,Te\kern -0.1em V}}}\xspace}
\newcommand{\gev}{\ensuremath{\mathrm{\,Ge\kern -0.1em V}}\xspace}
\newcommand{\mev}{\ensuremath{\mathrm{\,Me\kern -0.1em V}}\xspace}
\newcommand{\kev}{\ensuremath{\mathrm{\,ke\kern -0.1em V}}\xspace}
\newcommand{\kevcc}{\ensuremath{{\mathrm{\,ke\kern -0.1em V\!/}c^2}}\xspace}
\newcommand{\ev}{\ensuremath{\mathrm{\,e\kern -0.1em V}}\xspace}
\newcommand{\gevc}{\ensuremath{{\mathrm{\,Ge\kern -0.1em V\!/}c}}\xspace}
\newcommand{\mevc}{\ensuremath{{\mathrm{\,Me\kern -0.1em V\!/}c}}\xspace}
\newcommand{\gevcc}{\ensuremath{{\mathrm{\,Ge\kern -0.1em V\!/}c^2}}\xspace}
\newcommand{\gevgevcccc}{\ensuremath{{\mathrm{\,Ge\kern -0.1em V^2\!/}c^4}}\xspace}
\newcommand{\mevcc}{\ensuremath{{\mathrm{\,Me\kern -0.1em V\!/}c^2}}\xspace}

\def\mum  {\ensuremath{{\,\upmu\mathrm{m}}}\xspace}

\def\invfb   {\ensuremath{\mbox{\,fb}^{-1}}\xspace}

\newcommand{\chisq}{\ensuremath{\chi^2}\xspace}

\newcommand{\chisqip}{\ensuremath{\chi^2_{\text{IP}}}\xspace}

\def\gsim{{~\raise.15em\hbox{$>$}\kern-.85em
          \lower.35em\hbox{$\sim$}~}\xspace}
\def\lsim{{~\raise.15em\hbox{$<$}\kern-.85em
          \lower.35em\hbox{$\sim$}~}\xspace}

\newcommand{\Real}{\ensuremath{\mathcal{R}e}\xspace}

\def\ptot       {\mbox{$p$}\xspace}
\def\pt         {\mbox{$p_{\mathrm{ T}}$}\xspace}

\def\evtgen     {\mbox{\textsc{EvtGen}}\xspace}

\def\geant      {\mbox{\textsc{Geant4}}\xspace}

\def\photos     {\mbox{\textsc{Photos}}\xspace}

\def\pythia     {\mbox{\textsc{Pythia}}\xspace}

\def\roofit     {\mbox{\textsc{RooFit}}\xspace}

\def\tell1  {TELL1\xspace}
\def\ukl1   {UKL1\xspace}

\usepackage{cite} 
\usepackage{mciteplus}

\usepackage{longtable} 
\usepackage{multirow}

\begin{document}

\renewcommand{\thefootnote}{\fnsymbol{footnote}}
\setcounter{footnote}{1}


\begin{titlepage}
\pagenumbering{roman}

\vspace*{-1.5cm}
\centerline{\large EUROPEAN ORGANIZATION FOR NUCLEAR RESEARCH (CERN)}
\vspace*{1.5cm}
\noindent
\begin{tabular*}{\linewidth}{lc@{\extracolsep{\fill}}r@{\extracolsep{0pt}}}
\ifthenelse{\boolean{pdflatex}}
{\vspace*{-2.7cm}\mbox{\!\!\!\includegraphics[width=.14\textwidth]{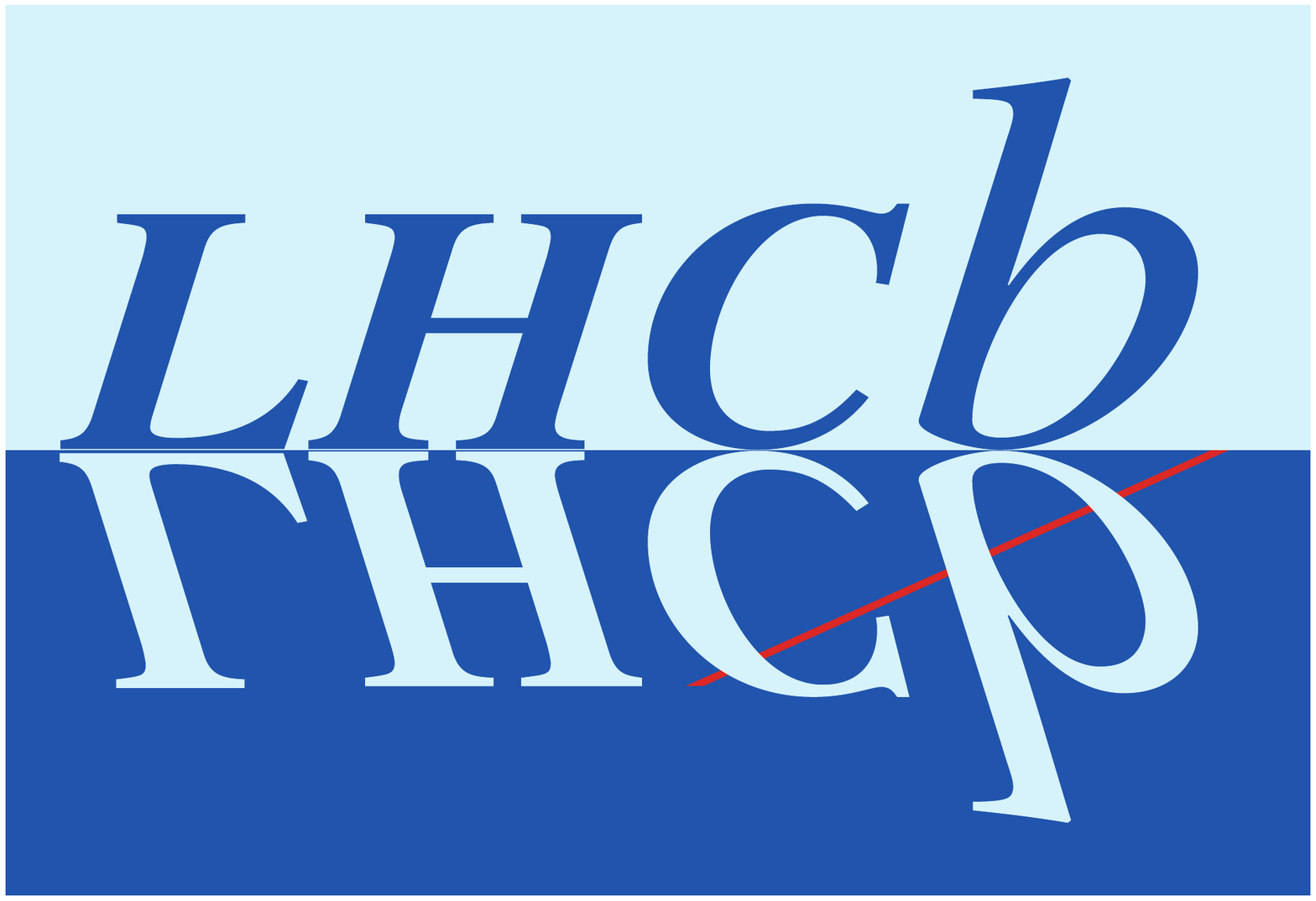}} & &}%
{\vspace*{-1.2cm}\mbox{\!\!\!\includegraphics[width=.12\textwidth]{lhcb-logo.eps}} & &}%
\\
 & & CERN-EP-2017-026 \\  
 & & LHCb-PAPER-2016-056 \\  
 & & \today \\ 
 & & \\
\end{tabular*}

\vspace*{2.0cm}

{\normalfont\bfseries\boldmath\huge
\begin{center}
Observation of the decay \\ $\Bs \to \etac \phi$ and evidence \\ for $\Bs \to \etac \pip\pim $ 
\end{center}
}

\vspace*{2.0cm}

\begin{center}
The LHCb collaboration\footnote{Authors are listed at the end of this paper.}
\end{center}
\vspace{\fill}

\begin{abstract}
  \noindent
A study of $\Bs \to \etac \phi$ and $\Bs \to \etac \pip\pim$ decays is performed using $pp$ collision data corresponding to an integrated luminosity of~3.0\invfb, collected with the LHCb detector in Run~1 of the LHC.  
The observation of the decay $\Bs \to \etac \phi$ is reported, where the \etac meson is reconstructed in the $p\bar p$, $K^+K^-\pi^+\pi^-$, $\pi^+\pi^-\pi^+\pi^-$ and $K^+K^-K^+K^-$ decay modes and the $\phi(1020)$ in the $K^+ K^-$ decay mode.
The decay $\Bs \to \jpsi \phi$ is used as a normalisation channel.
Evidence is also reported for the decay $\Bs \to \etac \pip\pim$, where the \etac meson is reconstructed in the $p\bar p$ decay mode, using the decay $\Bs \to \jpsi \pip\pim$ as a normalisation channel.
The measured branching fractions are
\begin{eqnarray*}
{\mathcal B (B^{0}_{s} \to \eta_{c} \phi)} &=& \left(5.01 \pm 0.53 \pm 0.27 \pm 0.63 \right) \times 10^{-4} \,, \nonumber \\                
 {\mathcal B (B^{0}_{s} \to \eta_{c} \pi^+ \pi^-)} &=& \left(1.76 \pm 0.59 \pm 0.12 \pm 0.29 \right) \times 10^{-4} \,,
\end{eqnarray*}
where in each case the first uncertainty is statistical, the second systematic and the third uncertainty is due to the limited knowledge of the external branching fractions.
\end{abstract}
\vspace*{1.0cm}

\begin{center}
  Published in JHEP 07 (2017) 021
\end{center}

\vspace{\fill}

{\footnotesize 
\centerline{\copyright~CERN on behalf of the \lhcb collaboration, licence \href{http://creativecommons.org/licenses/by/4.0/}{CC-BY-4.0}.}}
\vspace*{2mm}
\end{titlepage}


\newpage
\setcounter{page}{2}
\mbox{~}
\cleardoublepage


\renewcommand{\thefootnote}{\arabic{footnote}}
\setcounter{footnote}{0}


\pagestyle{plain} 
\setcounter{page}{1}
\pagenumbering{arabic}

%

\section{Introduction}
\label{sec:Introduction}

When a \Bs meson decays through the $\bar{b} \to \bar{c} c \bar{s}$ process, interference between the direct decay amplitude, and the amplitude after $\Bs-\Bsb$ oscillation, gives rise to a \CP-violating phase, \phis.
This phase is well predicted within the Standard Model (SM)~\cite{CKMfitter} and is sensitive to possible contributions from physics beyond the SM~\cite{Altmannshofer:2009ne,Altmannshofer:2007cs,Buras:2009if,Chiang:2009ev}.
The \phis phase is best measured using the ``golden'' channel\footnote{The simplified notation $\phi$ and $\etac$ are used to refer to the $\phi(1020)$ and the $\etac(1S)$ mesons throughout this article.} $\Bs \to \jpsi \phi$~\cite{LHCb-PAPER-2014-059, Abazov:2011ry, CDF:2011af, Khachatryan:2015nza, Aad:2016tdj} and the precision of this measurement is expected to be dominated by its statistical uncertainty until the end of LHC running. 
In addition to $\Bs \to \jpsi \phi$, other modes have been used to constrain \phis: \decay{\Bs}{\jpsi \pip \pim}~\cite{LHCb-PAPER-2014-059}, \decay{\Bs}{\Dsp\Dsm}~\cite{LHCb-PAPER-2014-051}, and \decay{\Bs}{\psi({\rm 2S})\phi}~\cite{LHCb-PAPER-2016-027}.  

\begin{figure}[h!]
\centering
\includegraphics[scale=0.35]{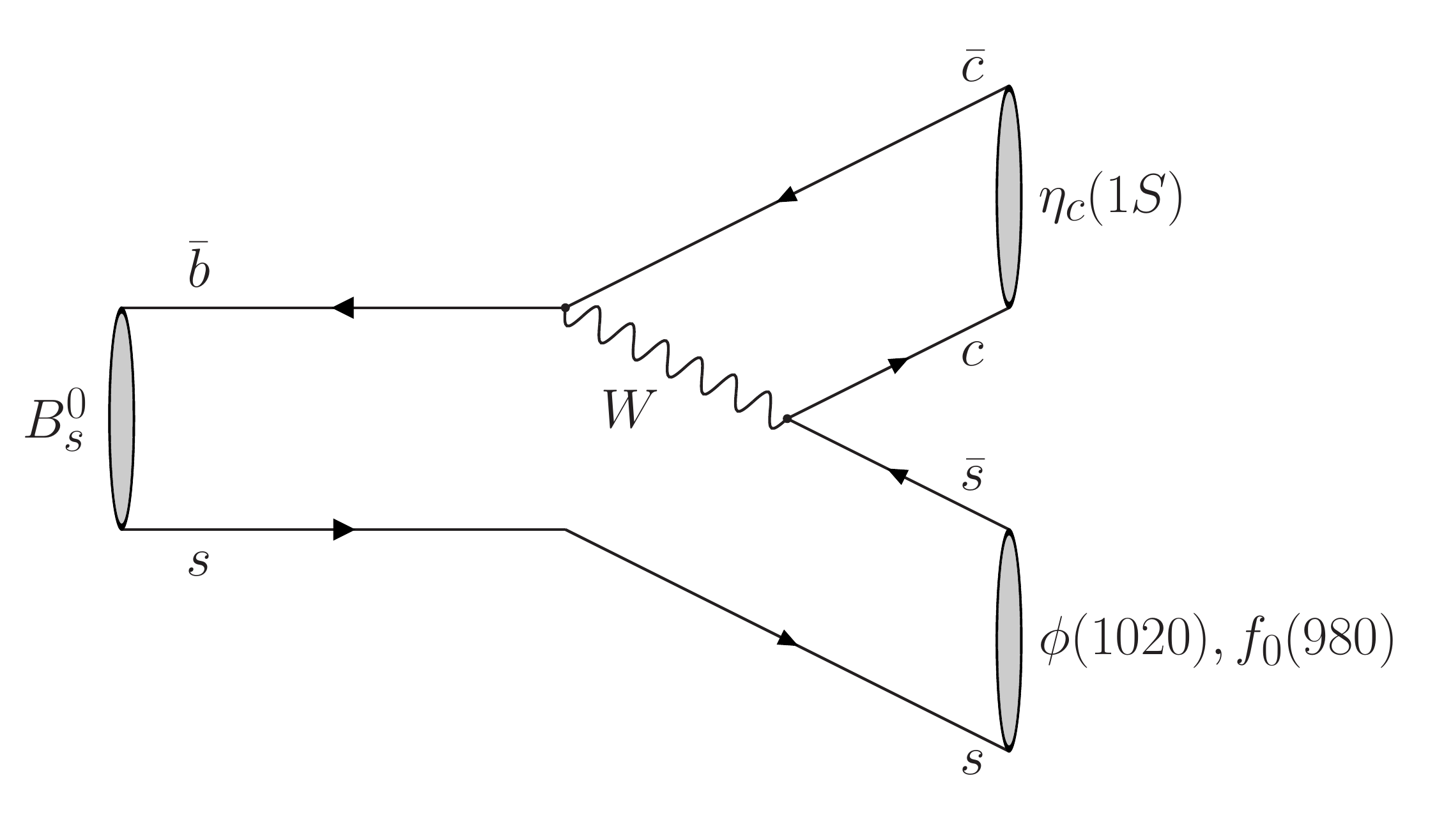}\\
\caption{Leading diagram corresponding to $\Bs \to \etac \phi$ and $\Bs \to \etac \pip\pim$ decays, where the $\pip\pim$ pair may arise from the decay of the $f_0(980)$ resonance.}
\label{fig:fig1}
\end{figure}

In this paper, the first study of $\Bs \to \eta_{c} \phi$ and $\Bs \to \eta_{c} \pip\pim$ decays is presented.\footnote{The use of charge-conjugate modes is implied throughout this article.} These decays also proceed dominantly through a $\bar b \to \bar c c \bar s$ tree diagram as shown in Fig.~\ref{fig:fig1}.
Unlike in \BsToJPsiPhi decays, the $\etac \phi$ final state is purely \CP-even, so that no angular analysis is required to measure the mixing phase $\phi_{s}$. 
However, the size of the data sample recorded by the LHCb experiment in LHC Run~1 is not sufficient to perform time-dependent analyses of $\Bs \to \eta_{c} \phi$ and $\Bs \to \eta_{c} \pip\pim$ decays. 
Instead, the first measurement of their branching fractions is performed.
No prediction is available for either $\BR(\Bs\to\etac\phi)$ or $\BR(\Bs\to\etac\pip\pim)$. 
Assuming
\begin{equation}
 \frac{\mathcal{B}(\Bs \rightarrow \etac \phi)}{\mathcal{B}(\Bs \rightarrow \jpsi \phi)}= 
\frac{\mathcal{B}(\Bd \rightarrow \etac K^0)}{\mathcal{B}(\Bd \rightarrow \jpsi K^0)} =
\frac{\mathcal{B}(\Bs \rightarrow \etac \pip\pim)}{\mathcal{B}(\Bs \rightarrow \jpsi \pip\pim)} \, 
\label{eq:br_estimate}
\end{equation}
allows $\BR(\Bs\to\etac\phi)$ and $\BR(\Bs\to\etac\pip\pim)$ to be estimated.
From the known values of $\BR(\Bd\to\etac\Kz)$, $\BR(\Bd\to\jpsi\Kz)$, $\BR(\Bs\to\jpsi\phi)$ and $\BR(\Bs\to\jpsi\pip\pim)$~\cite{PDG2016}, one finds 
\begin{eqnarray}\label{eq:br_estimate_phi}
\BR(\Bs\to\etac\phi) & = & \mathcal{O}(10^{-3})\,,  \\
\label{eq:br_estimate_pipi}
\BR(\Bs\to\etac\pip\pim) & = & \mathcal{O}(10^{-4}) \,.
\end{eqnarray}
The measurements presented in this paper are performed using a dataset corresponding to 3\invfb of integrated luminosity collected by the LHCb experiment in $pp$ collisions during 2011 and 2012 at centre-of-mass energies of 7\tev and 8\tev, respectively. 
The paper is organised as follows: Section~\ref{sec:Detector} describes the LHCb detector and the procedure used to generate simulated events; 
an overview of the strategy for the measurements of $\BR(\Bs\to\etac\phi)$ and $\BR(\Bs\to\etac\pip\pim)$ is given in Sec.~\ref{sec:strategy}; 
the selection of candidate signal decays is described in Sec.~\ref{sec:eventSelection};
the methods to determine the reconstruction and selection efficiencies are discussed in Sec.~\ref{sec:effi}.
Section~\ref{sec:fitModel} describes the fit models. 
The results and associated systematic uncertainties are discussed in Secs.~\ref{sec:results} and~\ref{sec:systematics}.
Finally, conclusions are presented in Sec.~\ref{sec:conclusions}.

%
\section{Detector and simulation}
\label{sec:Detector}

The \lhcb detector~\cite{Alves:2008zz,LHCb-DP-2014-002} is a single-arm forward spectrometer covering the \mbox{pseudorapidity} range $2<\eta <5$, designed for the study of particles containing \bquark or \cquark quarks. 
The detector includes a high-precision tracking system consisting of a silicon-strip vertex detector surrounding the $pp$ interaction region, a large-area silicon-strip detector located upstream of a dipole magnet with a bending power of about $4{\mathrm{\,Tm}}$, and three stations of silicon-strip detectors and straw drift tubes placed downstream of the magnet.
The tracking system provides a measurement of momentum, \ptot, of charged particles with a relative uncertainty that varies from 0.5\% at low momentum to 1.0\% at 200\gevc.
The minimum distance of a track to a primary vertex (PV), the impact parameter (IP), is measured with a resolution of $(15+29/\pt)\mum$, where \pt is the component of the momentum transverse to the beam, in\,\gevc.
Different types of charged hadrons are distinguished using information from two ring-imaging Cherenkov detectors. 
Photons, electrons and hadrons are identified by a calorimeter system consisting of scintillating-pad and preshower detectors, an electromagnetic calorimeter and a hadronic calorimeter. 
Muons are identified by a system composed of alternating layers of iron and multiwire proportional chambers. 

The online event selection is performed by a trigger~\cite{LHCb-DP-2012-004}, which consists of a hardware stage, based on information from the calorimeter and muon systems, followed by a software stage, which applies a full event reconstruction.

Samples of simulated events are used to determine the effects of the detector geometry, trigger, and selection criteria on the invariant-mass distributions of interest for this paper.
In the simulation, $pp$ collisions are generated using \pythia~\cite{Sjostrand:2006za,*Sjostrand:2007gs} with a specific \lhcb configuration~\cite{LHCb-PROC-2010-056}. 
The decay of the \Bs meson is described by \evtgen~\cite{Lange:2001uf}, which generates final-state radiation using \photos~\cite{Golonka:2005pn}. 
The interaction of the generated particles with the detector, and its response, are implemented using the \geant toolkit~\cite{Allison:2006ve, *Agostinelli:2002hh} as described in Ref.~\cite{LHCb-PROC-2011-006}.
Data-driven corrections are applied to the simulation to account for the small level of mismodelling of the particle identification (PID) performance~\cite{LHCb-DP-2012-003}.
In the simulation the reconstructed momentum of every track is smeared by a small amount in order to better match the mass resolution of the data.

%
\section{Analysis strategy}
\label{sec:strategy}

In the analysis of $\Bs\to\etac\phi$ decays, the $\phi$ meson is reconstructed in the $\Kp\Km$ final state and the \etac meson is reconstructed in the $p\bar p$, $K^+K^-\pi^+\pi^-$, $\pi^+\pi^-\pi^+\pi^-$ and $K^+K^-K^+K^-$ final states. 
For clarity, the three four-body final states are referred to as $4h$ throughout the paper.  
In determining the branching fraction, the decay $\Bs \to \jpsi \phi$ is used as a normalisation channel, where the $\jpsi$ meson is reconstructed in the same decay modes as the \etac meson.
A similar strategy is adopted for the measurement of the branching fraction of $\Bs\to\etac\pip\pim$ decays. 
However, due to the higher expected level of combinatorial background compared to $\Bs\to\etac\phi$ decays, the \etac and \jpsi mesons are reconstructed only in the $p \bar p$ final state in the measurement of $\BF(\Bs\to\etac\pip\pim)$. 

In both analyses, a two-stage fit procedure is performed. 
In the first stage, unbinned extended maximum likelihood (UML) fits are performed to separate signal candidates from background contributions. 
For the $\Bs \to \etac(\to p\bar p) \pip\pim$ decay the fit is done to the $p\bar p \pip\pim$ mass distribution, while for the decays $\Bs \to \etac(\to p\bar p) \phi(\to \Kp\Km)$ and $\Bs \to \etac(\to 4h) \phi(\to \Kp\Km)$ it is made to the two-dimensional $ p\bar p \Kp\Km$ versus $\Kp\Km$ or $4h \Kp\Km$ versus $\Kp\Km$ mass distributions, respectively.
The likelihood function is 
\begin{equation}
\label{eq:likelihood}
  {\cal L}({\mathbf N}, {\mathbf a}) = \frac{e^{-\sum_j N_j}}{n!} \prod_{l=1}^n \left ( \sum_j N_j {\cal P}_j(m;\bold{a}) \right) ,
\end{equation}
where $j$ stands for the event species, $N_j$ is the corresponding yield and ${\bf N}$ is the vector of yields $N_j$, ${\bf a}$ is the vector of fitted parameters other than yields, $n$ is the total number of candidates in the sample, and ${\cal P}_j(m)$ is the probability density function (PDF) used to parametrise the set of invariant-mass distributions $m$ considered. 
The \roofit package\cite{Verkerke:2003ir} is used to construct the negative log-likelihood function (NLL), which is minimised using \textsc{Minuit}~\cite{James:1975dr}.
Using information from these fits, signal weights for each candidate, $\omega_l$, are obtained using the $_s{\cal P}lot$ technique~\cite{Pivk:2004ty}.  

In the second stage, for $\Bs \to p\bar{p} \pip\pim$ candidates a weighted UML fit is made to the $p\bar{p}$ invariant-mass spectrum, and weighted UML fits of the $p\bar{p}$ and the $4h$ invariant-mass spectra are done for $\Bs \to p\bar{p} \phi$ and $\Bs \to 4h \phi$ candidates, respectively, to disentangle $\etac$ and $\jpsi$ candidates from nonresonant (NR) and remaining background contributions, as described in Sec.~\ref{sec:fitModel}.
For the weighted fits, the NLL function is given by
\begin{equation}
\label{eq:likelihood_weights}
- {\rm ln} {\cal L}({\mathbf N}, {\mathbf a})= \zeta \sum_j N_j -  \zeta \sum_l \omega_l \, {\rm ln} \left(\sum_j N_j {\cal P}_j(m;\bold{a}) \right) + \ln(n!),
\end{equation}
where $\zeta = \sum_l \omega_l / \sum_l \omega_l^2$ ensures proper uncertainty estimates from the weighted likelihood fit~\cite{Xie:2009rka}.
For the observed numbers of $\eta_c$ and $\jpsi$ candidates in final state $f$, $N_{\etac,f}$ and $N_{\jpsi,f}$, the measured branching fraction is
\begin{equation}
\label{eq:br_expression_all}
\mathcal{B}(\Bs \to \etac X) = \frac{N_{\etac,f}}{N_{\jpsi,f}}\times\mathcal{B}(\Bs \to \jpsi X)\times\frac{\mathcal{B}(\jpsi \to f)}{\mathcal{B}(\etac \to f)}\times\frac{{\varepsilon(\jpsi)_f}}{{\varepsilon(\etac)_f}}\, ,
\end{equation}
where $X$ refers to either the $\phi$ meson or the $\pip\pim$ pair.
The branching fractions $\mathcal{B}(\Bs \rightarrow \jpsi \phi)$, $\mathcal{B}(\Bs \rightarrow \jpsi \pip\pim)$, $\mathcal{B}(\jpsi \rightarrow f)$ and $\mathcal{B}(\etac \rightarrow f)$ are taken from Ref.~\cite{PDG2016}, and the efficiency correction factors, $\varepsilon$, are obtained from simulation.
In order to maximise the sensitivity to $\mathcal{B}(\Bs \rightarrow \etac \phi)$, a simultaneous fit to the $p\bar{p}$ and $4h$ invariant-mass spectra is performed.

%
\section{Event selection}
\label{sec:eventSelection}

A common strategy for the event selection, comprising several stages, is adopted for all final states. 
First, online requirements are applied at the trigger level, followed by an initial offline selection in which relatively loose criteria are applied. 
Boosted decision trees (BDTs)~\cite{Breiman}, implemented using the \textsc{TMVA} software package~\cite{Hocker:2007ht}, are then used to further suppress the combinatorial background arising from random combinations of tracks originating from any PV. 
Finally, the requirements on the output of the BDTs and on the PID variables are simultaneously optimised for each final state, to maximise the statistical significance of the signal yields. 

At the hardware trigger stage, events are required to have a muon with high \pt or a
  hadron with high transverse energy in the calorimeters.
The software trigger requires a two-, three- or four-track secondary vertex (SV) with a significant displacement from any PV. 
At least one charged particle must have a large transverse momentum and be inconsistent with originating from a PV. 
A multivariate algorithm~\cite{BBDT} is used for the identification of secondary vertices consistent with the decay of a $b$ hadron into charged hadrons. 
In addition, for the $4h$ final states, an algorithm is used to identify inclusive $\phi \to K^{+} K^{-}$ production at a secondary vertex, without requiring a decay consistent with a $b$ hadron. 

In the initial stage of the offline selection, candidates for $\mbox{\ensuremath{\Bs \to p\bar{p}\pip\pim}}$ and $\mbox{\ensuremath{\Bs \to p\bar{p} \Kp\Km}}$ $(\Bs \to 4h \Kp\Km)$ decays are required to have four (six) good quality, high-\pt tracks consistent with coming from a vertex that is displaced from any PV in the event. 
Loose PID criteria are applied, requiring the tracks to be consistent with the types of hadrons corresponding to the respective final states.
In addition, the $\Bs$ candidates, formed by the combination of the final-state candidates, are required to originate from a PV by requiring a small angle between the $\Bs$ candidate momentum vector and the vector joining this PV and the $\Bs$ decay vertex, and a small \chisqip, which is defined as the difference in the vertex-fit \chisq of the considered PV reconstructed with and without the candidate.
When forming the $\Bs$ candidates for $\Bs \to p\bar{p}\pip\pim$ and $\Bs \to p\bar{p} \Kp\Km$ decays, the $p\bar p$ mass resolution is improved by performing a kinematic fit~\cite{Hulsbergen:2005pu} in which the \Bs candidate is constrained to originate from its associated PV (that with the smallest value of \chisqip\ for the \Bs), and its reconstructed invariant mass is constrained to be equal to the known value of the $B^{0}_{s}$ mass~\cite{PDG2016}. 
No significant improvement of the $4h$ mass resolution is observed for $\Bs \to 4h \Kp\Km$ decays.
In order to reduce the combinatorial background, a first BDT, based on kinematic and topological properties of the reconstructed tracks and candidates, is applied directly at the initial stage of the offline selection of candidate $\Bs \to 4h K^+K^-$ decays. 
It is trained with events from dedicated simulation samples as signal and data from the reconstructed high-mass sidebands of the \Bs candidates as background.

In the second step of the selection, the offline BDTs are applied. They are trained using the same strategy as that used for the training of the first BDT. 
The maximum distance of closest approach between final-state particles, the transverse momentum, and the $\chi^{2}_{\rm IP}$ of each reconstructed track, as well as the vertex-fit $\chi^{2}$ per degree of freedom, the $\chi^{2}_{\rm IP}$, and the pointing angle of the \Bs candidates are used as input to the BDT classifiers used to select candidate $\Bs \to p\bar{p}\pip\pim$ and $\Bs \to p\bar{p} \Kp\Km$ decays.
For the $p\bar p K^{+}K^{-}$ final state, the direction angle, the flight distance significance and the $\chi^{2}_{\rm IP}$ of the reconstructed \Bs candidate are also used as input to the BDT, while the \pt of the \Bs candidate is used for the $p\bar p \pi^{+}\pi^{-}$ final state. The difference in the choice of input variables for the 
$p \bar pK^{+}K^{-}$ and the $p \bar p \pi^{+} \pi^{-}$ final states is due to different PID requirements applied to pions and kaons in the first stage of the offline selection.
The optimised requirements on the BDT output and PID variables for $\Bs \to p\bar{p}\pip\pim$ $(\Bs \to p\bar{p}\Kp\Km)$ decays retain $\sim 45 \%$ ($40 \%$) of the signal and reject more than $99\%$ ($99\%$) of the combinatorial background, inside the mass-fit ranges defined in Sec.~\ref{sec:fitModel}.

Dedicated BDT classifiers are trained to select candidate $\Bs \to 4h \Kp\Km$ decays using the following set of input variables: the \pt and the IP with respect to the SV of all reconstructed tracks; the vertex-fit $\chi^2$ of the $\etac$ and $\phi$ candidates; the vertex-fit $\chi^2$, the \pt, the flight-distance significance with respect to the PV of the \Bs candidate, and the angle between the momentum and the vector joining the primary to the secondary vertex of the \Bs candidate.
The optimised requirements on the BDT output and PID variables, for each of the $4h$ modes, retain about $50\%$ of the signal and reject more than $99\%$ of the combinatorial background inside the mass-fit ranges defined in Sec.~\ref{sec:fitModel}.

From simulation, after all requirements for $\Bs \to 4h \Kp\Km$ decays, a significant contamination is expected from $\Bs \to \Dsp 3h$ decays, where the $\Dsp$ decays to $\phi \pip$ and $3h$ is any combination of three charged kaons and pions.
This background contribution has distributions similar to the signal in the $4hK^+K^-$ and $K^+K^-$ invariant-mass spectra, while its distribution in the $4h$ invariant-mass spectrum is not expected to exhibit any peaking structure. 
In order to reduce this background contamination, the absolute difference between the known value of the \Ds mass~\cite{PDG2016} and the reconstructed invariant mass of the system formed by the combination of the $\Pphi$ candidate and any signal candidate track consistent with a pion hypothesis is required to be $> 17\mevcc$. 
This requirement is optimised using the significance of $\Bs \to \jpsi \Kp\Km$ candidates with respect to background contributions. This significance is stable for cut values in the range $[9, 25]\,$MeV$/c^2$, with a maximum at 17$\,$MeV$/c^2$, which removes about $90\%$ of $\Bs \to \Dsp 3h$ decays, with no significant signal loss.

%
\section{Efficiency correction}
\label{sec:effi}

\begin{table}[b]
\caption{Ratio of efficiencies between the normalisation and signal channels for each final state.} 
\label{tab:effi_correction_factor}
\centering
\begin{tabular}{l | ccccc}
 &  $\Bs \to 2K2\pi \phi$ & $\Bs \to 4\pi \phi$ & $\Bs \to 4K \phi$ & $\Bs \to p\bar{p} \phi$ & $\Bs \to p\bar{p} \pip\pim$  \\ 
\hline
$\frac{\varepsilon(\jpsi)}{\varepsilon(\etac)}$ &  $1.047 \pm 0.011$         & $1.068 \pm 0.016$         & $0.962 \pm 0.028$         & $1.038 \pm 0.009$ & $1.004 \pm 0.015 $\\
\end{tabular}
\end{table}

The efficiency correction factors appearing in Eq.~\ref{eq:br_expression_all} are obtained from fully simulated events.
Since the signal and normalisation channels are selected based on the same requirements and have the same final-state particles with very similar kinematic distributions, the ratio between the efficiency correction factors for $\Bs \to \etac X$ and $\Bs \to \jpsi X$ decays are expected to be close to unity.
The efficiency correction factors include the geometrical acceptance of the LHCb detector, the reconstruction efficiency, the efficiency of the offline selection criteria, including the trigger and PID requirements.
The efficiencies of the PID requirements are obtained as a function of particle momentum and number of charged tracks in the event using dedicated data-driven calibration samples of pions, kaons, and protons~\cite{LHCb-PROC-2011-008}.
The overall efficiency is taken as the product of the geometrical acceptance of the LHCb detector, the reconstruction efficiency and the efficiency of the offline selection criteria.
In addition, corrections are applied to account for different lifetime values used in simulation with respect to the known values for the decay channels considered.
The effective lifetime for \Bs decays to $ \etac\phi$ $(\etac\pip\pim)$ final state, being purely \CP-even (\CP-odd), is obtained from the known value of the decay width of the light (heavy) \Bs state~\cite{HFAG}.
The effective lifetime of $\Bs \to \jpsi \phi$ ($\Bs \to \jpsi \pip\pim$) decays is taken from Ref.~\cite{HFAG}.
The lifetime correction is obtained after reweighting the signal and normalisation simulation samples.
The final efficiency correction factors, given in Table~\ref{tab:effi_correction_factor}, are found to be compatible to unity as expected.

%
\section{Fit models}
\label{sec:fitModel}

In this section the fit models used for the measurement of the branching fractions are described, first the model used for $\Bs\to\etac\pip\pim$ decays in Sec.~\ref{sec:etacpipi}, then the model used for $\Bs\to\etac\phi$ decays in Sec.~\ref{sec:etacphi}.

\subsection{\boldmath{Model for $\Bs\to\etac\pip\pim$ decays}}
\label{sec:etacpipi}

Candidates are fitted in two stages.
First, an extended UML fit to the $p\bar{p} \pi^{+} \pi^{-}$ invariant-mass spectrum is performed in the range $5150$--$5540$\mevcc, to discriminate $\Bs\to p\bar{p}\pip\pim$ events from combinatorial background, $B^{0}  \to p \bar p \pip \pim$ decays, and $B^{0} \to p \bar p K \pi$ decays, where the kaon is misidentified as a pion.
The $p \bar p \pip \pim$ mass distribution of $\Bs  \to p \bar p \pip \pim$ and $B^{0}  \to p \bar p \pip \pim$ candidates are described by Hypatia functions~\cite{Santos:2013gra}. 
Both Hypatia functions share common core resolution and tail parameters. The latter are fixed to values obtained from simulation.
The distribution of the misidentified $B^{0} \to p \bar p K \pi$ background is described by a Crystal Ball function~\cite{Skwarnicki:1986xj}, with mode, power-law tail, and core resolution parameters fixed to values obtained from simulation.
The combinatorial background is modelled using an exponential function.
The mode and the common core resolution parameters of the Hypatia functions and the slope of the exponential functions, as well as all the yields, are allowed to vary in the fit to data. 
Using the information from the fit to the $p\bar{p}\pip\pim$ spectrum, signal weights are then computed and the background components are subtracted using the $_s{\cal P}lot$ technique~\cite{Pivk:2004ty}. 
Correlations between the $p\bar{p}$ and $p\bar{p}\pip\pim$ invariant-mass spectra, for both signal and backgrounds, are found to be negligible.

Second, a UML fit to the weighted $p\bar{p}$ invariant-mass distribution is performed in the mass range $2900$--$3200$\mevcc.
In this region, three event categories are expected to populate the $p\bar{p}$ spectrum: the \etac and \jpsi resonances, as well as a possible contribution from nonresonant $\Bs \to (p\bar p)_{\rm NR} \pip\pim$ decays.
The $p\bar{p}$ mass distribution of \etac candidates is described by the convolution of the square of the modulus of a complex relativistic Breit-Wigner function (RBW) with constant width and a function describing resolution effects. 
The expression of the RBW function is taken as
\begin{equation}
\label{eq:rBW}
R_{\rm res}(m;m_{\rm res},\Gamma_{\rm res}) \propto \frac{1}{m^2_{\rm res} - m^2 - i m_{\rm res}\Gamma_{\rm res}},
\end{equation}
where $m_{\rm res}$ and $\Gamma_{\rm res}$ are the pole mass and the natural width, respectively, of the resonance. 
From simulation, in the mass range considered, the $p\bar p$ invariant-mass resolution is found to be a few \mevcc, while $\Gamma_\etac = 31.8\pm0.8\mevcc$~\cite{PDG2016}. 
Thus, the $p\bar{p}$ distribution of \etac candidates is expected to be dominated by the RBW, with only small effects on the total \etac lineshape from the resolution.
On the other hand, due to the small natural width of the \jpsi resonance~\cite{PDG2016}, the corresponding lineshape is assumed to be described to a very good approximation by the resolution function only.
For the \etac and \jpsi lineshapes, Hypatia functions are used to parametrise the resolution, with tail parameters that are fixed to values obtained from simulation. A single core resolution parameter, $\sigma_{\rm res}^{c\bar{c}}$, shared between these two functions, is free to vary in the fit to data.
The \etac pole mass and the mode of the Hypatia function describing the \jpsi lineshape, which can be approximated by the pole mass of the resonance, are also free to vary, while the $\eta_{c}$ natural width is constrained to its known value~\cite{PDG2016}.
The possible contribution from $\Bs \to (p\bar p)_{\rm NR} \pip\pim$ decays is parametrised by a constant.

The angular distributions of P- and S-waves are characterised by a linear combination of odd- and even-order Legendre polynomials, respectively.
In the case of a uniform acceptance, after integration over the helicity angles, the interference between the two waves vanishes. 
For a non-uniform acceptance, after integration, only residual effects from the interference between $\etac(\to p\bar{p})\pip\pim$ and $\jpsi(\to p\bar{p})\pip\pim$ amplitudes can arise in the $p\bar{p}$ invariant mass spectra. 
Due to the limited size of the current data sample, these effects are assumed to be negligible.
Also, given the sample size and the small expected contribution of the NR $p\bar p$ component, interference between the $\etac(\to p\bar{p})\pip\pim$ and $(p\bar{p})_{\rm NR}\pip\pim$ amplitudes is neglected.

In order to fully exploit the correlation between the yields of \etac and \jpsi candidates, the former is parametrised in the fit, rearranging Eq.~\eqref{eq:br_expression_all}, as
\begin{equation}
N_\etac = N_\jpsi \times \frac{\mathcal{B}(\Bs \rightarrow \etac \pip\pim)}{\mathcal{B}(\Bs \rightarrow \jpsi \pip\pim)}  \times \frac{ \mathcal{B}(\etac \rightarrow p\bar{p}) }{ \mathcal{B}(\jpsi \rightarrow p\bar{p})} \times \frac{{\varepsilon(\etac)_{p\bar{p}}}}{{\varepsilon(\jpsi)_{p\bar{p}}}} \,,
\end{equation}
where $\BR(\Bs \to \etac \pip\pim)$ and $N_\jpsi$ are free parameters. 
The yield of the NR $p\bar p$ component is also free to vary.

\subsection{\boldmath{Model for $\Bs\to\etac\phi$ decays}}
\label{sec:etacphi}

The procedure and the fit model used to measure $\mathcal{B}( \Bs\to\etac\phi )$ is based on that described in Sec.~\ref{sec:etacpipi}.
However, several additional features are needed to describe the data, as detailed below.

The $\Kp\Km$ invariant mass is added as a second dimension in the first step fit, which here consists of a two-dimensional (2D) fit to the $p\bar{p}\Kp\Km$ or $4h\Kp\Km$ and $\Kp\Km$ invariant mass spectra.
This allows the contributions from $\phi \to \Kp\Km$ decays and nonresonant $\Kp\Km$ pairs to be separated.
Thus, the first step of the fitting procedure consists of four independent two-dimensional UML fits to the $p\bar{p}\Kp\Km$ versus $\Kp\Km$ and $4h\Kp\Km$ versus $\Kp\Km$ invariant-mass spectra in the ranges $5200$--$5500$\mevcc and $990$--$1050$\mevcc, respectively.\footnote{In order to better constrain the combinatorial background shape, the upper limit of the $p\bar{p}\Kp\Km$ invariant-mass range is extended to $5550$\mevcc.} 

Similar 2D fit models are used for each $4h$ mode. 
The $4h \Kp \Km$ distributions of $\Bs  \to 4h \phi$ signal and $\Bd  \to 4h \phi$ background contributions, as well as those of $\Bs  \to 4h \Kp\Km$ and $\Bd  \to 4h \Kp\Km$ backgrounds, are described by Hypatia functions.  
The $4h \Kp \Km$ distribution of the combinatorial background is parametrised using two exponential functions, one for when the $\Kp\Km$ pair arises from a random combination of two prompt kaons, and another for when the $\Kp\Km$ pair originates from the decay of a prompt $\phi$ meson.
The $K^+K^-$ distribution of each contribution including a $\phi$ in the final state is described by the square of the modulus of a RBW with mass-dependent width convolved with a Gaussian function accounting for resolution effects.
The $K^+K^-$ distributions of the contributions including a nonresonant $\Kp\Km$ pair are parametrised by linear functions.
The expression of the RBW with mass-dependent width describing the $\phi$ resonance is the analogue of Eq.~\eqref{eq:rBW}, with the mass-dependent width given by 
\begin{equation}
\label{equ:RBW_Width_formula}
\Gamma(m) = \Gamma_{\phi} \left( \frac{q}{q_{\phi}} \right)^{3} \left( \frac{m_{\phi}}{m}\right) X^2(qr),
\end{equation}
where $m_{\phi}=1019.461\pm0.019$\mevcc, $\Gamma_{\phi}=4.266\pm0.031$\mevcc~\cite{PDG2016}, and $q$ is the magnitude of the momentum of one of the $\phi$ decay products, evaluated in the resonance rest frame such that
\begin{equation}
\label{equ:q_formula}
q = \frac{1}{2}\sqrt{m^2-4m_{K^\pm}^2}. 
\end{equation}
with $m_{K^\pm} = 493.677 \pm 0.016\mevcc$~\cite{PDG2016}.
The symbol $q_{\phi}$ denotes the value of $q$ when $m = m_{\phi}$. The $X(qr)$ function is the Blatt-Weisskopf barrier factor~\cite{BlattWeisskopf:1952,VonHippel:1972fg} with a barrier radius of $r$. 
The value of the parameter $r$ is fixed at $3\unit{(GeV/c)^{-1}}$.
Defining the quantity $z = qr$, the Blatt-Weisskopf barrier function for a spin-1 resonance is given by
\begin{equation}
X(z) = \sqrt{\frac{1+z_{\phi}^2}{1+z^2}}, 
\end{equation}
where $z_{\phi}$ represents the value of $z$ when $m = m_{\phi}$.

The same 2D fit model is used for the $p\bar{p}$ mode with an additional component accounting for the presence of misidentified $B^{0} \to p \bar p K \pi$ background events. 
The $p\bar{p} \Kp \Km$ and $\Kp \Km$ distributions of $B^{0} \to p \bar p K \pi$ candidates are described by a Crystal Ball function and a linear function, respectively.
 
Using the sets of signal weights computed from the 2D fits, the $p\bar{p}$ and $4h$ spectra are obtained after subtraction of background candidates from $\Bd$ decays and $\Bs$ decays with nonresonant $\Kp\Km$ pairs as well as combinatorial background.
Correlations between the invariant-mass spectra used in the 2D fits and the $p\bar{p}$ or $4h$ spectrum are found to be negligible.
A simultaneous UML fit is then performed to the weighted $p\bar{p}$ and $4h$ invariant-mass distributions, with identical mass ranges of $2820$--$3170$\mevcc.
Different models are used to describe the $p\bar{p}$ and $4h$ spectra.

The $p\bar{p}$ invariant-mass spectrum is modelled similarly to the description in Sec.~\ref{sec:etacpipi}.
However, as shown in Sec.~\ref{sec:results}, the fit to the $p\bar{p}$ spectrum for $\Bs \to p\bar{p} \pip\pim$ decays yields a contribution of NR $p\bar{p}$ decays compatible with zero. 
Thus, here, the contribution of such decays is fixed to zero and only considered as a source of systematic uncertainty, as described in Sec.~\ref{sec:systematics}.

For the $4h$ modes, in addition to $\Bs \to \etac \phi$ and $\Bs \to \jpsi \phi$ decays, other contributions are expected in the mass range considered:
$\Bs \to 4h \phi$ decays, where the $4h$ system is in a nonresonant state with a total angular momentum equal to zero, and where \Bs decays proceed via intermediate resonant states decaying in turn into two or three particles for instance, $\Bs \to PP^\prime \phi$ decays, where $P$ and $P^\prime$ could be any resonance such as $\Kstar(892)$, $\rho(770)$, $\phi(1020)$, $\Pomega(782)$, $\Pf_2(1270)$, $\Pf'_2(1525)$ and $\Pa_2(1320)$.
Similarly to $\Bs \to \Ds3h$ decays, all these decays are expected to have smooth distributions in the $4h$ invariant-mass spectra. 
Therefore, lacking information from previous measurements, all these contributions are merged into one category, denoted $(4h)_{\rm bkg}$. The $4h$ nonresonant contribution is denoted $(4h)_{\rm NR}$.
The $\etac$ being a pseudoscalar particle, interference between $\Bs \to \etac(\to 4h) \phi$ and $\Bs \to (4h)_{\rm NR}\phi$ amplitudes for each $4h$ final state are accounted for in the model. 
On the other hand, given the large number of amplitudes contributing to the $(4h)_{\rm bkg}$ event category, the net effect of all interference terms is assumed to cancel.
Similarly to the $p\bar{p}$ fit model, terms describing residual effects of the interference between the \jpsi and the other fit components are neglected.
The total amplitude for each of the $4h$ modes, integrated over the helicity angles, is then given by
\begin{equation}
\label{eq:ampForm1}
\left|A(m_f;c^f_k,\mathbf{a})\right|^2 =   \sum_k \left|c^f_k {R}_k(m_f;\mathbf{a})\right|^2 + 2\Real( c^f_\etac {R}_\etac(m_f;\mathbf{a}) c^{f\ast}_{\rm NR} {R}^\ast_{\rm NR}(m_f;\mathbf{a}) )\, ,
\end{equation}
where $R_k(m_f;\mathbf{a})$ is the line-shape of the component $k$, $\mathbf{a}$ represents the line-shape parameters, $c_k^f$ are complex numbers such that $c^f_k = \alpha_k^f \,e^{i\varphi_k^f}$ where $\alpha_k^f$ and $\varphi_k^f$ are the magnitude and the strong phase of amplitude $k$, and $m_f$ is one of the $4h$ invariant masses. 
The $\etac$ and the $\jpsi$ resonances are described similarly to the $p\bar{p}$ mode, and the $(4h)_{\rm NR}$ and $(4h)_{\rm bkg}$ components are described using exponential functions.

Finally, taking into account the detector resolution, the total function, ${\mathcal F}_{\rm tot}$, used to describe the invariant-mass spectra $m_f$ is given by
\begin{align}
\begin{split}
{\mathcal F}_{\rm tot}(m_f;c^f_k,\mathbf{a},\mathbf{a}^{\prime})  &= \left|A(m_f;c^f_k,\mathbf{a})\right|^2  \otimes \mathcal{R}(\mathbf{a}^{\prime} (m_f)) \\
 				&=  \xi^f_{\etac}\frac{{\mathcal F}_{\etac}(m_f)}{\int_{m_f} {\mathcal F}_{\etac}(m_f) {\rm d}m_f} + \xi^f_{\jpsi}\frac{{\mathcal F}_{\jpsi}(m_f)}{\int_{m_f} {\mathcal F}_{\jpsi}(m_f) {\rm d}m_f} \label{eq:amplitudePDF} \\ 
 				&  \quad + \xi^f_{\rm NR}\frac{{\mathcal F}_{\rm NR}(m_f)}{\int_{m_f} {\mathcal F}_{\rm NR}(m_f) {\rm d}m_f} + \xi^f_{\rm bkg}\frac{{\mathcal F}_{\rm bkg}(m_f)}{\int_{m_f} {\mathcal F}_{\rm bkg}(m_f) {\rm d}m_f} \\
 				&  \quad + 2 \sqrt{\xi^f_{\etac}\xi^f_{\rm NR}}\frac{{\mathcal F}_{\rm I}(m_f)}{\int_{m_f} \sqrt{{\mathcal F}_{\etac}(m_f){\mathcal F}_{\rm NR}(m_f)} {\rm d}m_f} \,,	
\end{split}
\end{align}
with $\xi^f_k = (\alpha_k^f)^2$ and where the expressions for ${\mathcal F}_{k}(m_f)$ are
\begin{eqnarray}
{\mathcal F}_{\etac}(m_f)  &=&  \left|R_{\etac}(m_f;\mathbf{a})\right|^{2} \otimes \mathcal{R}(\mathbf{a}^{\prime} (m_f)) , \\
{\mathcal F}_{\jpsi}(m_f)  &=& \mathcal{R}(\mathbf{a}^{\prime} (m_f)) , \\ 
{\mathcal F}_{\rm NR}(m_f)  &=& e^{\kappa_{\rm NR} m_f} \otimes \mathcal{R}(\mathbf{a}^{\prime} (m_f)) , \\ 
{\mathcal F}_{\rm bkg}(m_f)  &=& e^{\kappa_{\rm bkg} m_f} \otimes \mathcal{R}(\mathbf{a}^{\prime} (m_f)) , \\
{\mathcal F}_{\rm I}(m_f)  &=& \big( e^{\frac{\kappa_{\rm NR} m_f}{2}}\Real\big[R_{\etac}(m_f;\mathbf{a})e^{i\delta\varphi}\big] \big) \otimes \mathcal{R}(\mathbf{a}^{\prime} (m_f)) , 
\end{eqnarray}
where $\delta\varphi$ is the difference between the strong phases of $(4h)_{\rm NR} \phi$ and $\etac(\to 4h) \phi$ amplitudes. 
The integrals in Eq.~\eqref{eq:amplitudePDF} are calculated over the mass range in which the fit is performed.
Only the \etac and \jpsi components are used in the expression for ${\mathcal F}_{\rm tot}(m_{p\bar p})$.
The fit fractions FF$_k$ measured for each component, as well as the interference fit fraction FF$_{\rm I}$ between the $\etac$ and the NR amplitudes for the $4h$ modes, are calculated as:
\begin{eqnarray}
\label{equ:fit_fractions_res}
\mathrm{FF}_k^f &=& \int_{m_f} \frac{\xi^f_{k} {\mathcal F}_{k}(m_f)}{{\mathcal F}_{tot}(m_f) \int_{m_f} {\mathcal F}_{k}(m_f) {\rm d}m_f} {\rm d}m_f\,, \\
\label{equ:fit_fractions_inter}
\mathrm{FF}^f_{\rm I} &=& \int_{m_f} \frac{2 \sqrt{\xi^f_{\etac}\xi^f_{\rm NR}} {\mathcal F}_{\rm I}(m_f)}{{\mathcal F}_{tot}(m_f)\int_{m_f} \sqrt{{\mathcal F}_{\etac}(m_f){\mathcal F}_{\rm NR}(m_f)} {\rm d}m_f} {\rm d}m_f\,.
\end{eqnarray}

The resolution, $\mathcal{R}(\mathbf{a}^{\prime} (m_f))$, is described by a Hypatia function, with parameters $\mathbf{a}^{\prime} (m_f)$ that depend on the final state and the invariant-mass region.
They are estimated using dedicated simulation samples in two mass regions: a high-mass region around the \jpsi resonance, and a low-mass region around the \etac resonance.

As in the model for $\Bs \to p\bar{p} \pip\pim$ decays, the branching fraction $\mathcal{B}(\Bs \to \etac \phi)$ is directly determined in the fit.
In this configuration, the squared magnitudes of the \etac amplitudes, $\xi^f_\etac$, are parametrised as 
\begin{equation}
\label{eq:BRetac4hbis}
	\xi^f_{\etac} = \xi^f_{\jpsi} \times \frac{\mathcal{B}(\Bs \rightarrow \etac \phi)}{\mathcal{B}(\Bs \rightarrow \jpsi \phi)}  \times \frac{ \mathcal{B}(\etac \rightarrow f) }{ \mathcal{B}(\jpsi \rightarrow f)} \times  \frac{{\varepsilon(\etac)_{f}}}{{\varepsilon(\jpsi)_{f}}} \,.
\end{equation}

In the simultaneous fit to the $p\bar{p}$ and $4h$ invariant-mass spectra several parameters are allowed to take different values depending on the final state: the intensities $\xi^f_k$ (free to vary), the slopes $\kappa_{\rm bkg}$ and $\kappa_{\rm NR}$ of the $(4h)_{\rm bkg}$ and $(4h)_{\rm NR}$ exponentials, respectively, (free to vary), the relative strong phase between the $(4h)_{\rm NR}$ and \etac amplitudes (free to vary) as well as the low and high mass resolution parameters (fixed).
The \etac pole mass, the mode of the Hypatia function describing the \jpsi and the branching fraction $\mathcal{B}(\Bs \rightarrow \etac \phi)$ are common parameters across all final states and are free to vary in the fit. 
The \etac width is fixed to the world average value taken from Ref.~\cite{PDG2016}.
For each mode, $\xi_{\jpsi}$ and $\varphi_\etac$ are fixed as reference to 1 and 0, respectively. 

%
\section{Results}
\label{sec:results}


\begin{table*}[b]
\begin{center}
\caption{\label{tab:results_1stStep_Fit} Yields of the different final states as obtained from the fit to the $p\bar{p}\pip\pim$ invariant-mass distribution and from the 2D fits in the $p\bar{p}(4h)\Kp\Km \times \Kp\Km$ invariant-mass planes. Only statistical uncertainties are reported. The abbreviation ``n/a'' stands for ``not applicable''.}
\setlength{\tabcolsep}{0.0pc}
\begin{tabular*}{\textwidth}{@{\extracolsep{\fill}}l | cccc}
  &\multicolumn{4}{c}{Yield}\\
 Mode & $\Bs \to {\rm Mode}$ & $\Bd \to {\rm Mode}$ & Combinatorial & $\Bd \to p\bar{p} \Kp\pim$  \\ 
\hline
$p\bar{p}\pip\pim$ & $179 \pm 32$  & $384 \pm 43$ & $3261 \pm 119$ & $897 \pm  69 \phantom{0}$ \\ 

\hline
\hline
$p\bar{p} \phi$    & $447 \pm 24$  & $ 13 \pm  7$  & $\phantom{0} 43 \pm 17$    & \multirow{2}{*}{$ 11      \pm 14$}\\ 
$p\bar{p} \Kp\Km$  & $\phantom{0} 10 \pm 11$  & $\!\! -4 \pm   5$ & $106 \pm 19$    & \\ 

\hline
$2K2\pi \phi$      & $586 \pm 34$  & $ \phantom{00}7 \pm 17$  & $419 \pm 39$   & n/a \\ 
$2K2\pi \Kp\Km\phantom{0}$    & $\phantom{0} 86 \pm 21$  & $\phantom{0} 18 \pm 16$ & $329 \pm 33$   & n/a \\ 

\hline
$4\pi \phi$        & $502 \pm 33$  & $\phantom{0} 77 \pm 23$ & $380 \pm 43$   & n/a \\ 
$4\pi \Kp\Km$      & $111 \pm 25$  & $\phantom{0} 67 \pm 24$ & $599 \pm 43$   & n/a \\ 

\hline
$4K \phi$          & $151 \pm 15$  & $\phantom{0} 6 \pm 5$ & $ \phantom{0} 44  \pm 13$  & n/a \\ 
$4K \Kp\Km$        & $ -3 \pm  4\,\,$  & $\!\!\!\! -10  \pm 9$ & $ \phantom{0} 44  \pm 11$  & n/a \\ 
\end{tabular*}
\end{center}
\end{table*}

The yields of the various decay modes determined by the UML fit to the $p\bar{p}\pip\pim$ invariant mass distribution, and from the 2D fits to the $p\bar{p}(4h)\Kp\Km$ versus $\Kp\Km$ invariant mass planes, are summarised in Table~\ref{tab:results_1stStep_Fit}.
The mass distributions and the fit projections are shown in Appendix~\ref{app:fitProjections}.
The $p\bar{p}\pip\pim$ and 2D fit models are validated using large samples of pseudoexperiments, from which no significant bias is observed.

\begin{figure}[t]
\centering
\includegraphics[height=6.0cm]{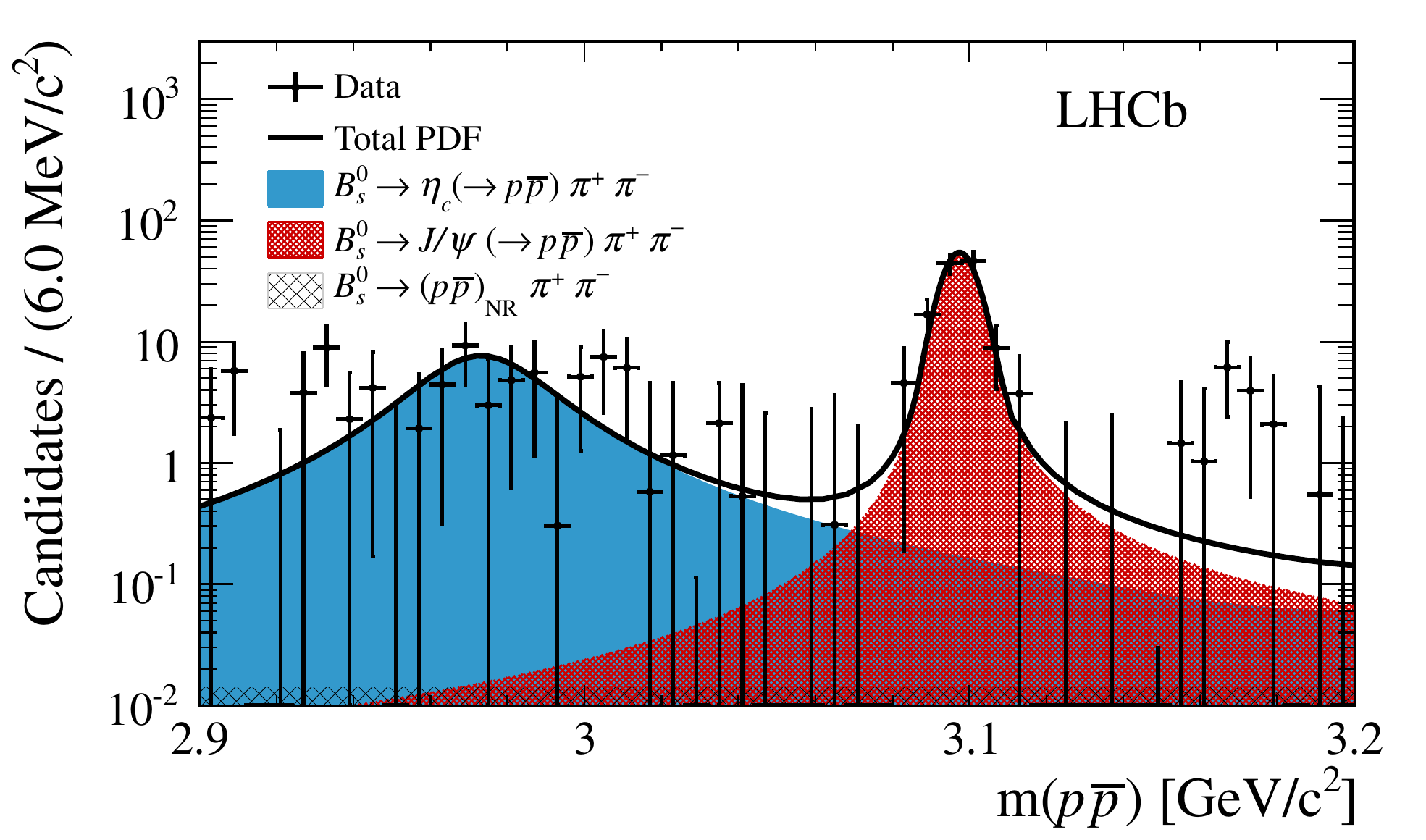}
\caption{\label{fig:fig2} Distribution of $p\bar p$ invariant-mass for $\Bs\to p\bar{p}\pip\pim$ candidates obtained by the $_s{\cal P}lot$ technique. The solid black curve is the projection of the total fit result. The full blue, tight-cross-hatched red and wide-cross-hatched black histograms show the \etac, \jpsi and nonresonant $p\bar p$ contributions, respectively. The structure visible around $3.15\,$GeV is found to be consistant with a statistical fluctuation.}
\end{figure}
\begin{figure}[t]
\centering
\includegraphics[width=9.0cm]{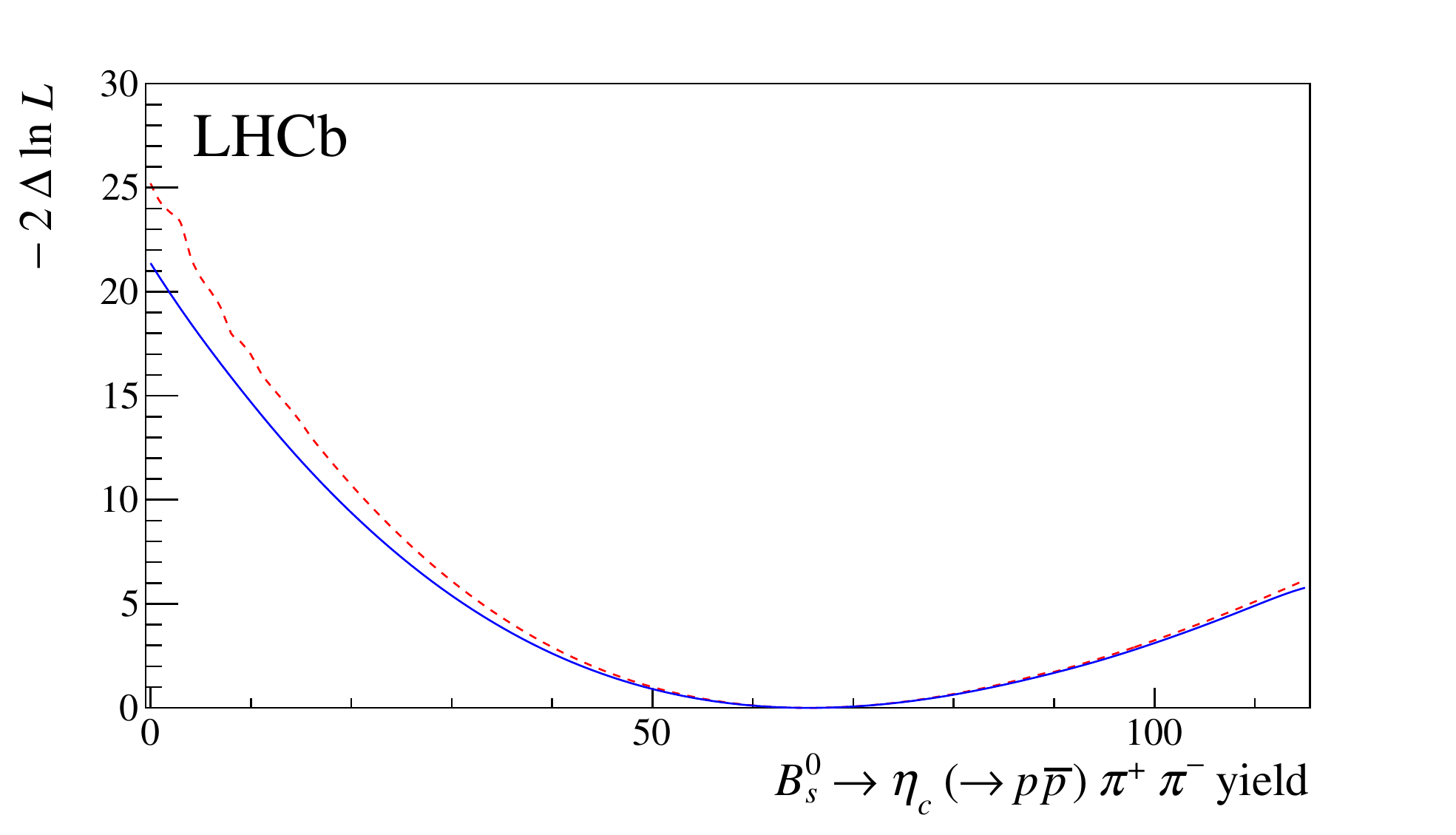}
\caption{\label{fig:fig3} One-dimensional scan of $-2 \Delta \ln L$ as a function of the \etac yield in the fit to the $p\bar{p}$ invariant-mass distribution for $\Bs \to p\bar{p}\pip\pim$ candidates. 
The dotted red and solid blue curves correspond to the result of the scan including statistical only and summed (statistical and systematic) uncertainties, respectively.}
\end{figure}

The $p\bar{p}$ invariant-mass distribution for $\Bs\to p\bar{p}\pip\pim$ candidates, and the projection of the fit are shown in Fig.~\ref{fig:fig2}. 
The values of the \etac and \jpsi shape parameters as well as the yields are given in Table~\ref{tab:etacPiPifit}. 
The branching fraction for the $\Bs\to \etac \pip\pim$ decay mode is found to be
\begin{equation}
\label{eq:etacPiPiBR}
 {\mathcal B (B^{0}_{s} \to \eta_{c} \pi^+\pi^-)} =  (1.76 \pm 0.59 \pm 0.12 \pm 0.29) \times 10^{-4} \,,
\end{equation}
 where the two first uncertainties are statistical and systematic,
 respectively, and the third uncertainty is due to the limited
 knowledge of the external branching fractions. The systematic
 uncertainties on the branching fraction are discussed in
 Sec.~\ref{sec:systematics}. The significance of the presence of
 $\Bs\to \etac \pip\pim$ decays in the $p\bar{p}$ invariant-mass
 spectrum is estimated, as $\sqrt{-2 \Delta \ln L}$, from the
   difference between the log-likelihood ($\ln L$) values for $N_\etac = 0$ and the value
   of $N_\etac$ that minimises $\ln L$.
For the estimation of the significance, $N_\etac$ is not parametrised as a function of $\BF(\Bs \to \etac \pip\pim)$, but is a free parameter in the fit.
As shown in Fig.~\ref{fig:fig3}, the significance of the \etac component in the fit to the $p\bar{p}$ invariant-mass distribution is $5.0$ standard deviations ($\sigma$) with statistical uncertainties and $4.6\sigma$ when including systematic uncertainties.
The latter is obtained by adding Gaussian constraints to the likelihood function.
This result is the first evidence for $\Bs\to \etac \pip\pim$ decays.
\begin{table}[b!!!]
\begin{center}
\caption{ \label{tab:etacPiPifit} Results of the fit to the $p\bar p$ invariant-mass spectra weighted for $\Bs\to p\bar{p}\pip\pim$ candidates. Uncertainties are statistical only. The parameter $N_{\rm NR}$ corresponds to the yield of $B^0_s \to (p\bar p)_{\rm NR} \pi^+ \pi^-$ candidates. The \etac yield does not appear since it is parametrised as a function of $\BF(\Bs\to \etac \pip\pim)$, the measured value of which is reported in Eq.~\eqref{eq:etacPiPiBR}.}
\setlength{\tabcolsep}{0.0pc}
\begin{tabular}{c|c}
\hline
$\quad m_{\etac} $ ($\!$\mevcc) $\quad$ &$\quad$ $\!\! 2973 \pm  8$$\quad$ \\
$m_{\jpsi}$ ($\!$\mevcc) &$\quad$ $3096.9  \pm 1.0$ $\quad$\\
$\sigma_{\rm res}^{c\bar{c}}$ ($\!$\mevcc) & $\quad$$\phantom{000}4.8  \pm  0.8$$\quad$\\
\hline
$N_\jpsi$ & $\quad$$\phantom{00}113  \pm  48$$\quad$ \\ 
$N_{\rm NR}$ & $\quad$$\phantom{000}0.5  \pm  8.5$$\quad$ \\
\hline
\end{tabular}
\end{center}
\end{table}

\begin{figure}[t]
\hspace{-15pt}
\begin{tabular}{cc}
\includegraphics[height=5.5cm]{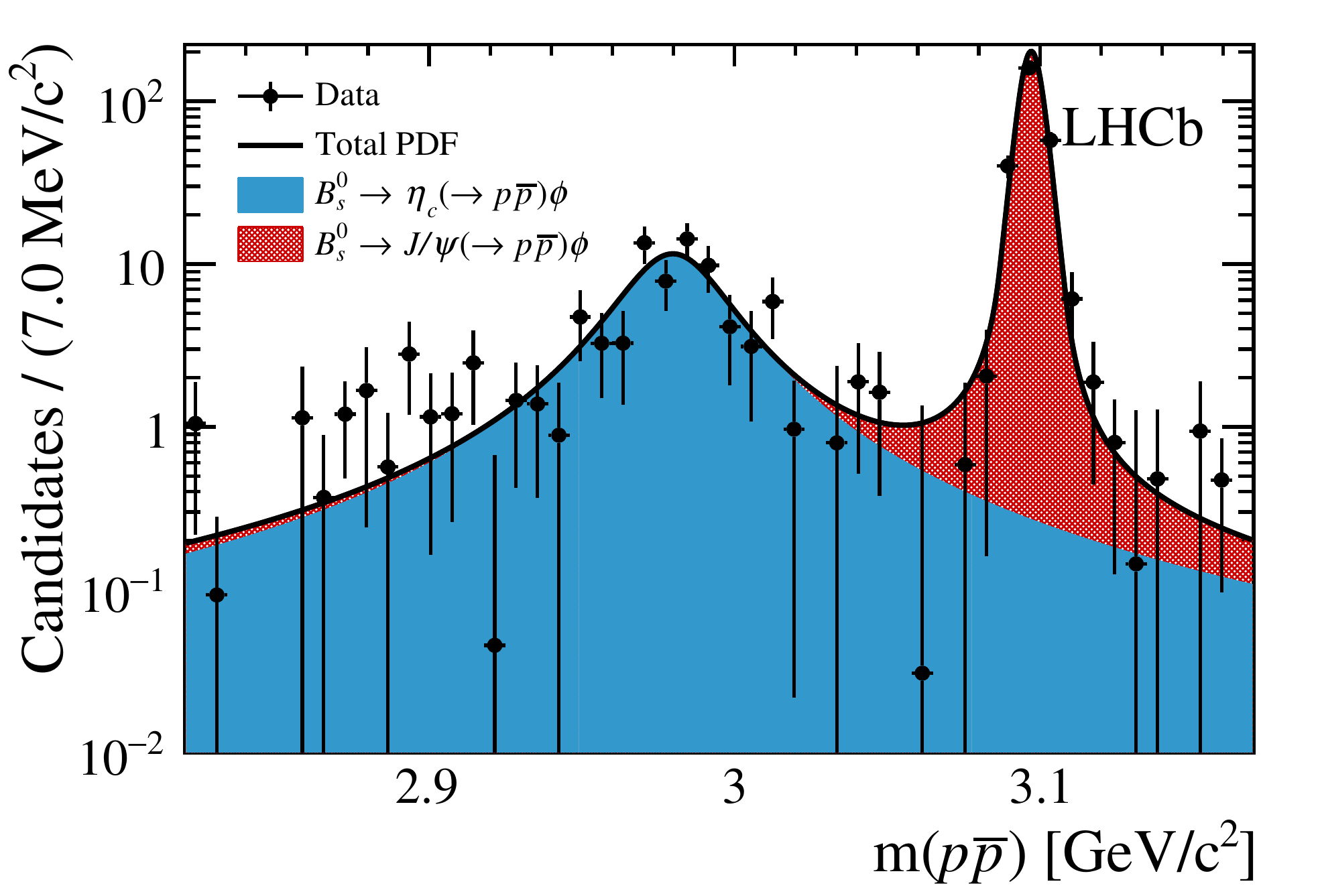} & 
\includegraphics[height=5.5cm]{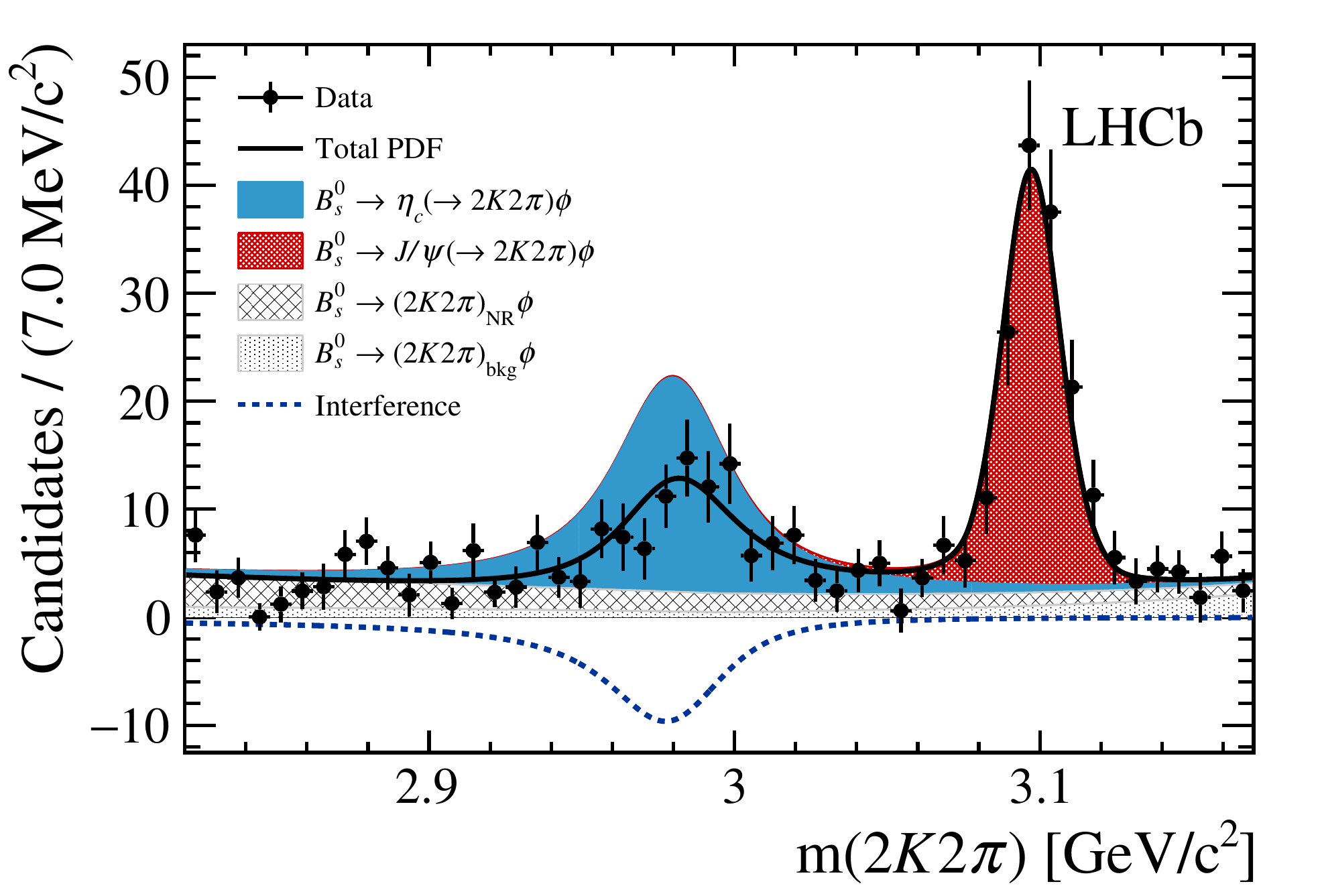}\\
\includegraphics[height=5.5cm]{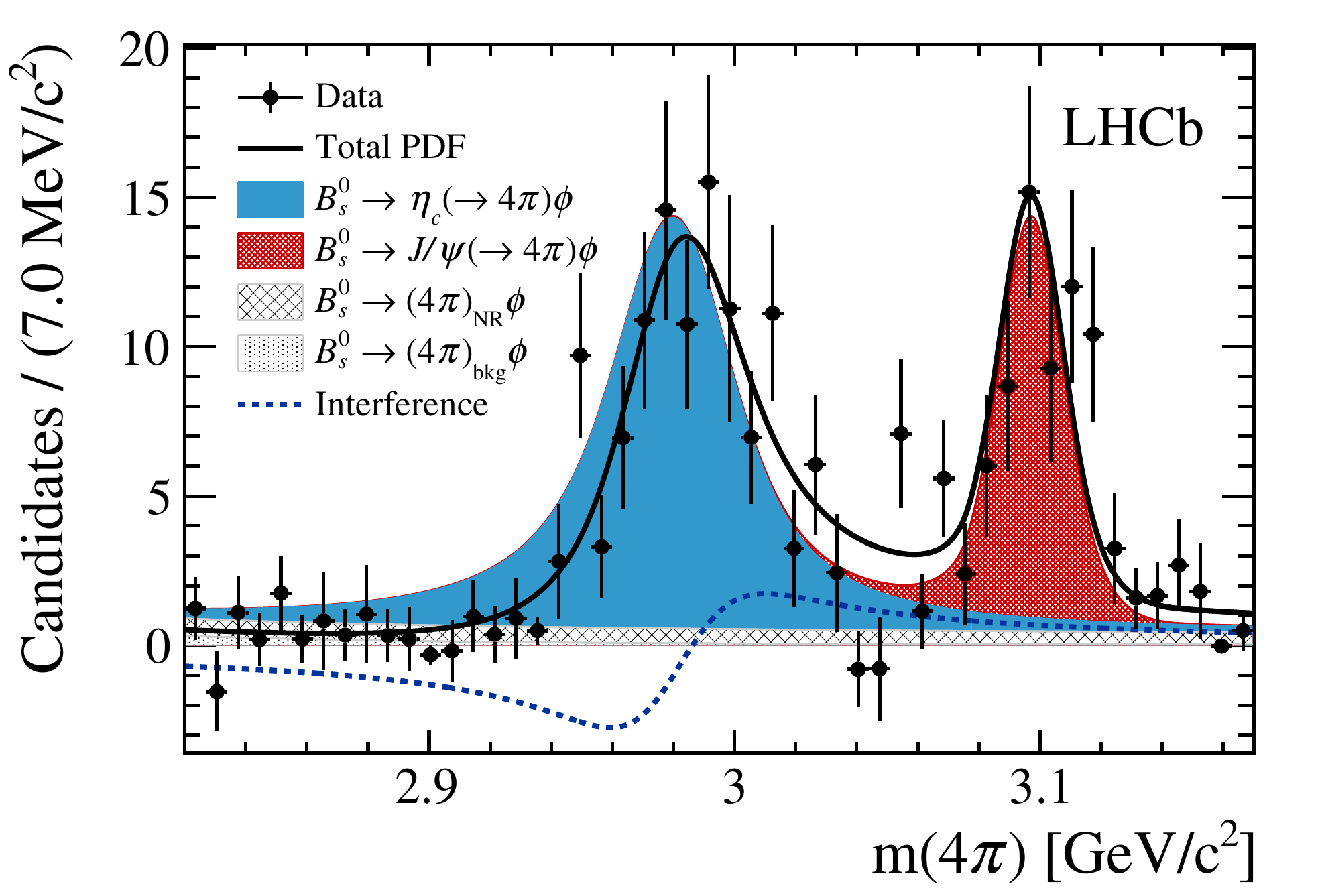} & 
\includegraphics[height=5.5cm]{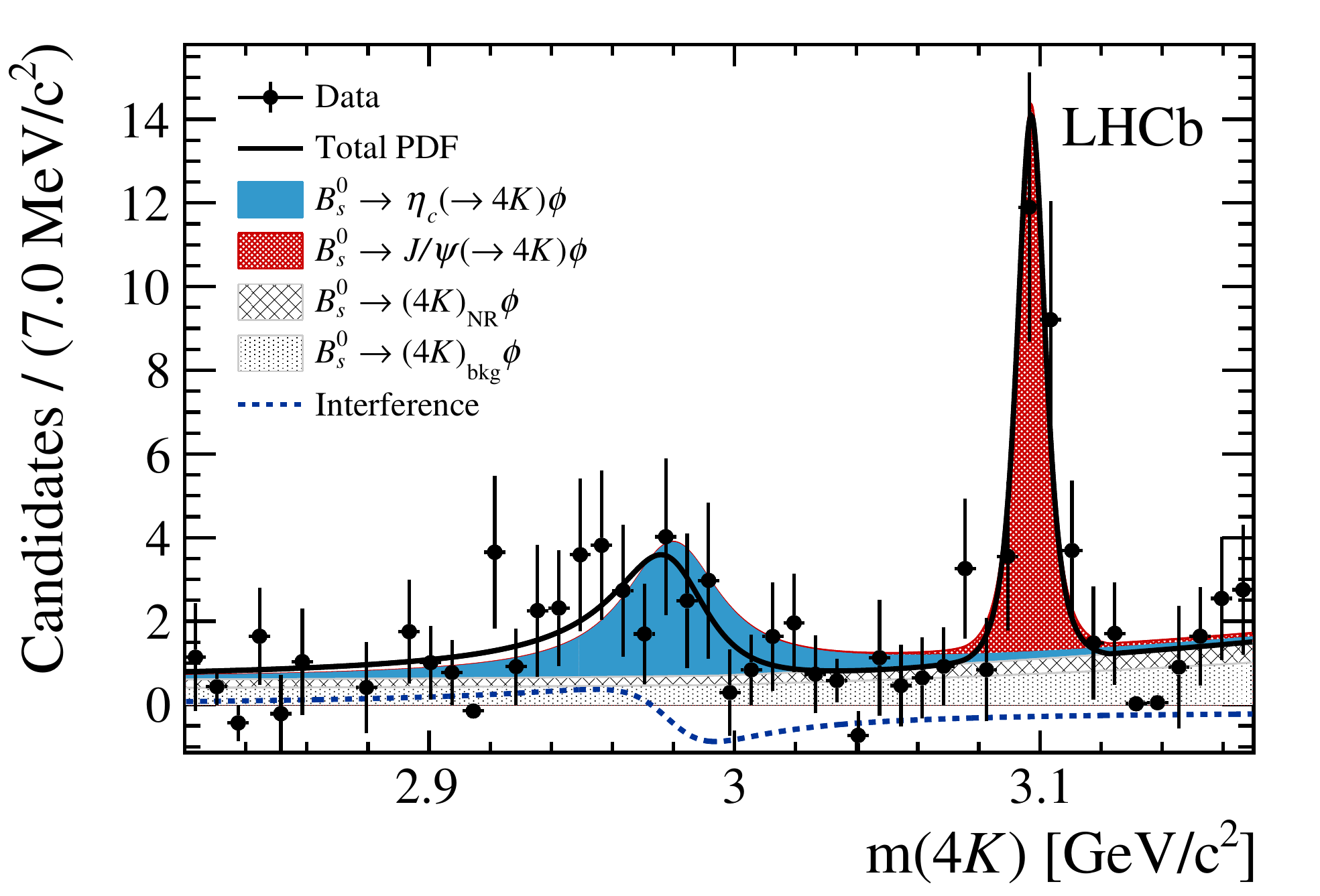}\\
\end{tabular}
\caption{\label{fig:fig4} Distribution of the $p\bar p$ and each of the $4h$ invariant-mass spectra for $\Bs\to p\bar{p}(4h)\phi$ candidates obtained by the $_s{\cal P}lot$ technique. 
The solid black curve is the total result of the simultaneous fit. The full blue and tight-cross-hatched red histograms show the \etac and \jpsi contributions, respectively. In addition, for each of the $4h$ modes, the wide-cross-hatched and dotted black histograms show the $4h_{\rm NR}$ and $4h_{\rm bkg}$ contributions, respectively. The dashed blue curves show the interference between $\Bs \to \etac(\to 4h)\phi$ and $\Bs \to (4h)_{\rm NR}\phi$ amplitudes.}
\end{figure}
\clearpage
The $p\bar{p}$ and $4h$ invariant-mass distributions for $\Bs\to p\bar{p}\phi$ and $\Bs\to 4h\phi$ candidates, and the projection of the simultaneous fit are shown in Fig.~\ref{fig:fig4}. 
The values of the shape parameters, of the magnitudes and of the relative strong phases are given in Table~\ref{tab:etacPhifit}. 
The statistical correlation matrix of the simultaneous fit is given in Appendix~\ref{app:corrMatrix}. 
The fit fractions are given in Table~\ref{tab:simAmpFitResult_FFs}.
The measured branching fraction for the $\Bs\to \etac \phi$ decay mode is
\begin{equation}
\label{eq:etacPhiBR}
 {\mathcal B (B^{0}_{s} \to \eta_{c} \phi)} = \left(5.01 \pm 0.53 \pm 0.27 \pm 0.63\right) \times 10^{-4} \,,                
\end{equation}
where the two first uncertainties are statistical and systematic, respectively, and the third uncertainty is due to the limited knowledge of the external branching fractions.
This measurement corresponds to the first observation of $\Bs\to \etac \phi$ decays.
As a cross-check, individual fits to the $p\bar{p}$ and to each of the $4h$ invariant-mass spectra give compatible values of $\mathcal B (B^{0}_{s} \to \eta_{c} \phi)$ within statistical uncertainties.
The precision of the $\mathcal B (B^{0}_{s} \to \eta_{c} \phi)$
measurement obtained using each of the $4h$ modes is limited compared to the $p\bar{p}$ mode.
This is expected due to the presence of additional components below the \etac and \jpsi resonance in the $4h$ invariant-mass spectra, and due to the interference between $\Bs \to \etac(\to 4h)\phi$ and $\Bs \to (4h)_{\rm NR}\phi$ amplitudes.
The measurement of $\mathcal B (B^{0}_{s} \to \eta_{c} \phi)$ from the simultaneous fit is largely dominated by the $p\bar{p}$ mode.

\begin{table}[h]
\caption{Result of the simultaneous fit to the $p\bar{p}$ and $4h$ invariant-mass spectra. Uncertainties are statistical only. The \jpsi and \etac magnitudes do not appear since they are set to unity as reference and parametrised as a function of $\BF(\Bs\to \etac \phi)$, respectively. In the simultaneous fit, the $m_{\etac}$ and $m_{\jpsi}$ parameters are shared across the four modes.
The measured value of $\BF(\Bs\to \etac \phi)$ is reported in Eq.~\eqref{eq:etacPhiBR}. The abbreviation ``n/a'' stands for ``not applicable''.}
\label{tab:etacPhifit}
\centering
\begin{tabular}{c | cccc } 
Parameter &\multicolumn{4}{c}{Mode}\\ 
 \hline 
& $2K2 \pi$ & $4 \pi$ & $4K$ & $p \bar{p}$  \\ 
 \hline 
$m_{\etac}$ ($\!$\mevcc) & \multicolumn{4}{c}{$2980.0 \pm 2.3$}\\ 
$m_{\jpsi}$ ($\!$\mevcc) & \multicolumn{4}{c}{$3097.2 \pm 0.2$} \\ 
\hline 
$\kappa_{\textrm{NR}}$ ($\!$\mevcc) & $-4 \pm 4$ & $-1 \pm 8$ & $5 \pm 4$ & n/a\\ 
$\kappa_{\textrm{bkg}}$ ($\!$\mevcc) & \phantom{-}$13 \pm 8$ & $-14 \pm 18$ & $4.9 \pm 2.9$ & n/a\\ 
\hline 
$\xi_{\textrm{NR}}$ ($\!$\mevcc) & \phantom{-0}$0.62 \pm 0.29$ & \phantom{-0}$0.42 \pm 0.31$ & $0.5 \pm 0.6$ & n/a\\ 
$\xi_{\textrm{bkg}}$ ($\!$\mevcc) & \phantom{-0}$0.31 \pm 0.25$ & \phantom{-0}$0.09 \pm 0.11$ & $1.1 \pm 0.7$ & n/a\\ 
\hline 
$\delta\varphi$ (rad) & \phantom{-0}$1.73 \pm 0.18$ & \phantom{-0}$2.9 \pm 0.6$ & $0.3 \pm 0.9$ & n/a\\ 
\end{tabular} 
\end{table}
\begin{table}[h]
\caption{Fit fractions obtained from the parameters of the simultaneous fit to the $p\bar{p}$ and $4h$ invariant-mass spectra. Uncertainties are statistical only. Due to interference between $\Bs \to \etac(\to 4h) \phi$ and $\Bs \to (4h)_{\rm NR} \phi$ amplitudes, for the $4h$ final states the sum of fit fractions, $\sum_k {\rm FF}_k $, may be different from unity. The abbreviation ``n/a'' stands for ``not applicable''.}
\label{tab:simAmpFitResult_FFs}
\centering
\begin{tabular}{l | rrrc } 
& \multicolumn{1}{c}{$2K2\pi$} & \multicolumn{1}{c}{$4\pi$} & \multicolumn{1}{c}{$4K$} & $p\bar{p}$  \\ 
 \hline 
  FF$_{\jpsi}$                                 & $ 0.39 \pm 0.03 \phantom{0}$  & $ 0.28 \pm 0.03 \phantom{0}$  & $ 0.29 \pm 0.05 \phantom{0}$ & $0.76 \pm 0.03$   \\ 
  FF$_{\etac}$                                 & $ 0.49 \pm 0.05 \phantom{0}$  & $ 0.63 \pm 0.07 \phantom{0}$  & $ 0.31 \pm 0.03 \phantom{0}$ & $0.24 \pm 0.03$  \\ 
  FF$_{\textrm{bkg}}$                           & $ 0.12 \pm 0.10 \phantom{0}$  & $ 0.02 \pm 0.03 \phantom{0}$  & $ 0.32 \pm 0.19 \phantom{0}$ & n/a   \\ 
  FF$_{\textrm{NR}}$                       & $ 0.24 \pm 0.11 \phantom{0}$  & $ 0.12 \pm 0.09 \phantom{0}$  & $ 0.15 \pm 0.16 \phantom{0}$ & n/a  \\ 
 \hline 
  $\sum_k {\rm FF}_k $                         & $ 1.24 \pm 0.07 \phantom{0}$  & $ 1.05 \pm 0.11 \phantom{0}$  & $ 1.08 \pm 0.08 \phantom{0}$ & $1.00$  \\
\end{tabular} 
\end{table}

%
\clearpage
\section{Systematic uncertainties}
\label{sec:systematics}

As the expressions for $\mathcal{B}(\Bs \rightarrow \etac \pip\pim)$ and $\mathcal{B}(\Bs \rightarrow \etac \phi)$ are based on the ratios of observed quantities, only sources of systematic uncertainties inducing different biases to the number of observed \etac and \jpsi candidates are considered. 
The dominant source of systematic uncertainties is due to the knowledge of the external branching fractions. 
These are estimated by adding Gaussian constraints on the external branching fractions in the fits, with widths corresponding to their known uncertainties~\cite{PDG2016}. 
A summary of the systematic uncertainties can be found in Table~\ref{tab:systsummary}. 

To assign systematic uncertainties due to fixing of PDF parameters, the fits are repeated by varying all of them simultaneously. 
The resolution parameters, estimated from simulation, are varied according to normal distributions, taking into account the correlations between the parameters and with variances related to the size of the simulated samples.
The external parameters are varied within a normal distribution of mean and width fixed to their known values and uncertainties~\cite{PDG2016}.
This procedure is repeated 1000 times, and for each iteration a new value of the branching fraction is obtained.
The systematic uncertainties on the branching fraction are taken from the variance of the corresponding distributions. 

\begin{table}[b]
\centering
\caption{Summary of systematic uncertainties. The ``Sum'' of systematic uncertainties is obtained from the quadratic sum of the individual sources, except the external branching fractions, which are quoted separately.
All values are in $\%$ of the measured branching fractions. The abbreviation ``n/a'' stands for ``not applicable''.}
\label{tab:systsummary}
\begin{tabular}{c c c }
Source     &  \multicolumn{2}{c}{Value [\%]} \\ 
& $\mathcal{B}(\Bs \rightarrow \etac \pip\pim)$ & $\mathcal{B}(\Bs \rightarrow \etac \phi)$ \\
\hline
Fixed PDF parameters            & $5.7$ & $1.4$ \\
Efficiencies                    & $3.4$ & $0.8$ \\
Fit bias                        & $1.7$ & $1.4$ \\
Resolution model                & $0.6$ & $4.4$ \\
$\phi(1020)$ barrier radius     & n/a   & $1.6$\\ 
Acceptance $(4h)$               & n/a   & $1.6$\\
Nonresonant $p\bar p$          & n/a   & $1.0$\\
\hline
Sum                             & $6.8$ & $5.4$\\
\hline
External branching fractions      & $16.4$ & $12.6$ \\
\end{tabular}
\end{table}

The systematic uncertainty due to the fixing of the values of the efficiencies is estimated by adding Gaussian constraints to the likelihood functions, with widths that are taken from the uncertainties quoted in Table~\ref{tab:effi_correction_factor}.

The presence of intrinsic biases in the fit models is studied using parametric simulation. 
For this study, 1000 pseudoexperiments are generated and fitted using the nominal PDFs, where the generated parameter values correspond to those obtained in the fits to data. 
The biases on the branching fractions are then calculated as the difference between the generated values and the mean of the distribution of the fitted branching fraction values. 

To assign a systematic uncertainty from the model used to describe the detector resolution, the fits are repeated for each step replacing the Hypatia functions by bifurcated Crystal Ball functions, the parameters of which are obtained from simulation. 
The difference from the nominal branching fraction result is assigned as a systematic uncertainty. 

The Blatt-Weisskopf parameter $r$ of the $\phi$ is arbitrarily set to $3\unit{(\gevc)^{-1}}$.
To assign a systematic uncertainty due to the fixed value of this $r$ parameter, the fits are repeated for different values taken in the range $1.5$--$5.0\unit{(\gevc)^{-1}}$.
The maximum differences from the nominal branching fraction result are assigned as systematic uncertainties.

To assign a systematic uncertainty due to the assumption of a uniform acceptance, the simultaneous fit is repeated after correcting the $4h$ invariant-mass distributions for acceptance effects. 
A histogram describing the acceptance effects in each of the $4h$ invariant-mass spectra is constructed from the ratio of the normalised $4h$ invariant-mass distributions taken from simulated samples of $\Bs \to (4h) \phi$ phase space decays, obtained either directly from \evtgen, or after processing through the full simulation chain.     
The simultaneous fit is repeated after applying weights for each event from the central value of its bin in the $4h$ invariant-mass distribution.
The difference from the nominal branching fraction result is assigned as a systematic uncertainty. 
No significant dependence on the binning choice was observed.

The systematic uncertainty due to neglecting the presence of a nonresonant $p\bar{p}$ contribution in the $p\bar{p}$ spectrum for $\Bs \to p\bar{p}\phi$ candidates is estimated by repeating the simultaneous fit with an additional component described by an exponential function, where the slope and the yield are allowed to vary.
The difference from the nominal branching fraction result is assigned as a systematic uncertainty.

\section{Conclusions}
\label{sec:conclusions}

This paper reports the observation of $\Bs \to \etac \phi$ decays and the first evidence for $\Bs  \to \etac \pip\pim$ decays. 
The branching fractions are measured to be
\begin{eqnarray*}
{\mathcal B (B^{0}_{s} \to \eta_{c} \phi)} &=& \left(5.01 \pm 0.53 \pm 0.27 \pm 0.63 \right) \times 10^{-4} \,, \nonumber \\                
 {\mathcal B (B^{0}_{s} \to \eta_{c} \pi^+ \pi^-)} &=& \left(1.76 \pm 0.59 \pm 0.12 \pm 0.29 \right) \times 10^{-4} \,,
\end{eqnarray*}
where in each case the two first uncertainties are statistical and systematic, respectively, and the third uncertainties are due to the limited knowledge of the external branching fractions.
The significance of the $\Bs \to \etac \pip\pim$ decay mode, including systematic uncertainties, is $4.6\sigma$.
The results for $\BF(\Bs  \to \etac \pip\pim)$ and $\BF(\Bs  \to \etac \phi)$ are in agreement with expectations based on Eqs.~\eqref{eq:br_estimate},~\eqref{eq:br_estimate_phi} and~\eqref{eq:br_estimate_pipi}.

The data sample recorded by the LHCb experiment in Run~1 of the LHC is not sufficiently large to allow a measurement of the \CP-violating phase \phis from time-dependent analysis of $\Bs \to \etac \phi$ or $\Bs  \to \etac \pip\pim$ decays.
However, in the future with significant improvement of the hadronic trigger efficiencies~\cite{LHCb-TDR-016}, these decay modes may become of interest to add sensitivity to the measurement of \phis. 

%

\section*{Acknowledgements}


\noindent We express our gratitude to our colleagues in the CERN
accelerator departments for the excellent performance of the LHC. We
thank the technical and administrative staff at the LHCb
institutes. We acknowledge support from CERN and from the national
agencies: CAPES, CNPq, FAPERJ and FINEP (Brazil); MOST and NSFC (China);
CNRS/IN2P3 (France); BMBF, DFG and MPG (Germany); INFN (Italy);
FOM and NWO (The Netherlands); MNiSW and NCN (Poland); MEN/IFA (Romania);
MinES and FASO (Russia); MinECo (Spain); SNSF and SER (Switzerland);
NASU (Ukraine); STFC (United Kingdom); NSF (USA).
We acknowledge the computing resources that are provided by CERN, IN2P3 (France), KIT and DESY (Germany), INFN (Italy), SURF (The Netherlands), PIC (Spain), GridPP (United Kingdom), RRCKI and Yandex LLC (Russia), CSCS (Switzerland), IFIN-HH (Romania), CBPF (Brazil), PL-GRID (Poland) and OSC (USA). We are indebted to the communities behind the multiple open
source software packages on which we depend.
Individual groups or members have received support from AvH Foundation (Germany),
EPLANET, Marie Sk\l{}odowska-Curie Actions and ERC (European Union),
Conseil G\'{e}n\'{e}ral de Haute-Savoie, Labex ENIGMASS and OCEVU,
R\'{e}gion Auvergne (France), RFBR and Yandex LLC (Russia), GVA, XuntaGal and GENCAT (Spain), Herchel Smith Fund, The Royal Society, Royal Commission for the Exhibition of 1851 and the Leverhulme Trust (United Kingdom).

\newpage

{\noindent\normalfont\bfseries\Large Appendix}
\appendix
\section{Fit projections}
\label{app:fitProjections} 

\begin{figure}[h]
\centering
\includegraphics[scale=0.75]{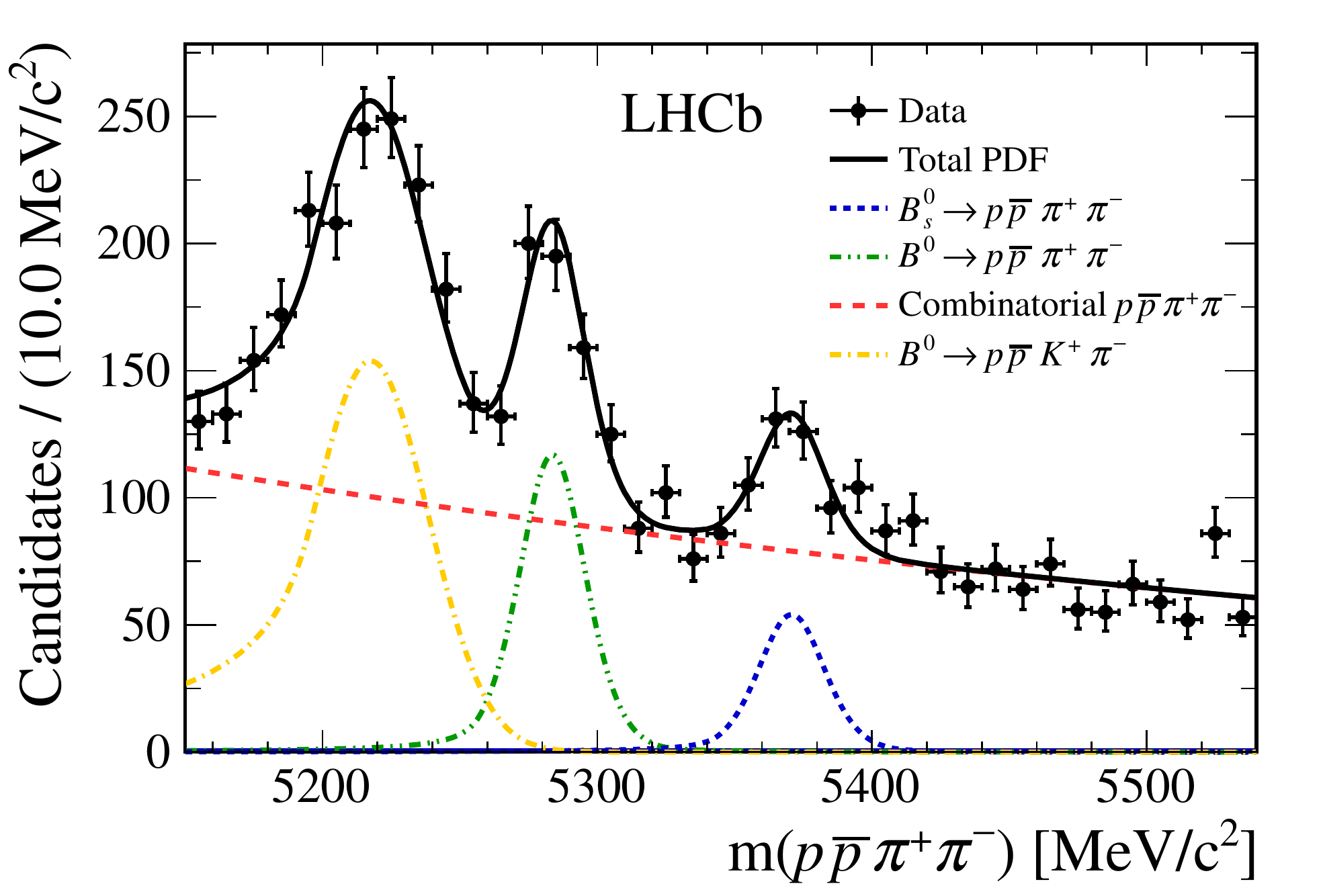}
\caption{\label{fig:fig5} 
Distribution of the $p\bar{p}\pip\pim$ invariant mass. 
Points with error bars show the data.  
The solid curve is the projection of the total fit result. 
The short-dashed blue, the dashed-double-dotted green, the dashed-single-dotted yellow and medium-dashed red curves show the $\Bs \to p\bar{p}\pip\pim$, $\Bd \to p\bar{p}\pip\pim$, $\Bd \to p\bar{p}\Kp\pim$ and combinatorial background contributions, respectively.
}
\end{figure}

\begin{figure}[htbp]
\centering
\includegraphics[width=15.cm]{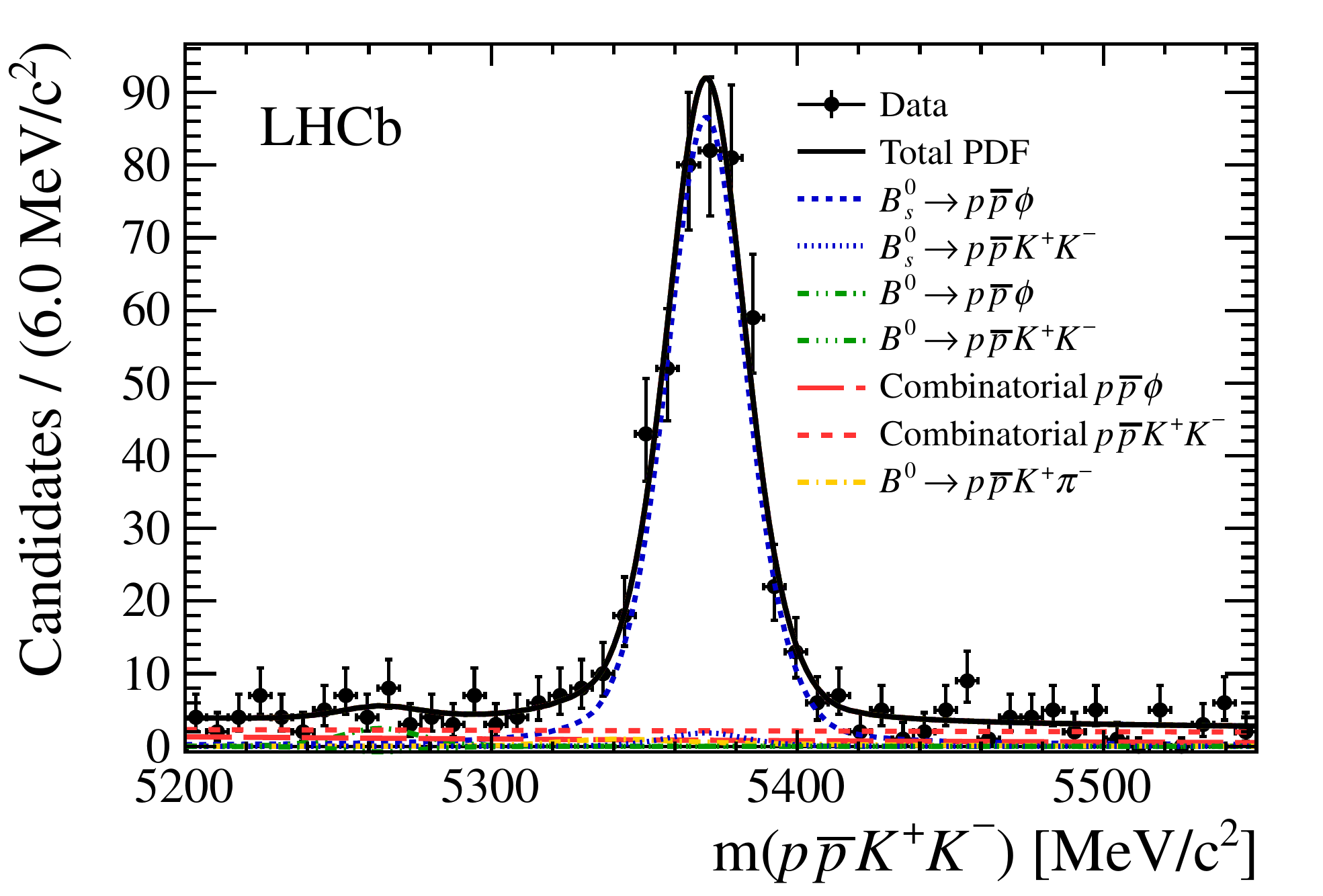}
\includegraphics[width=15.cm]{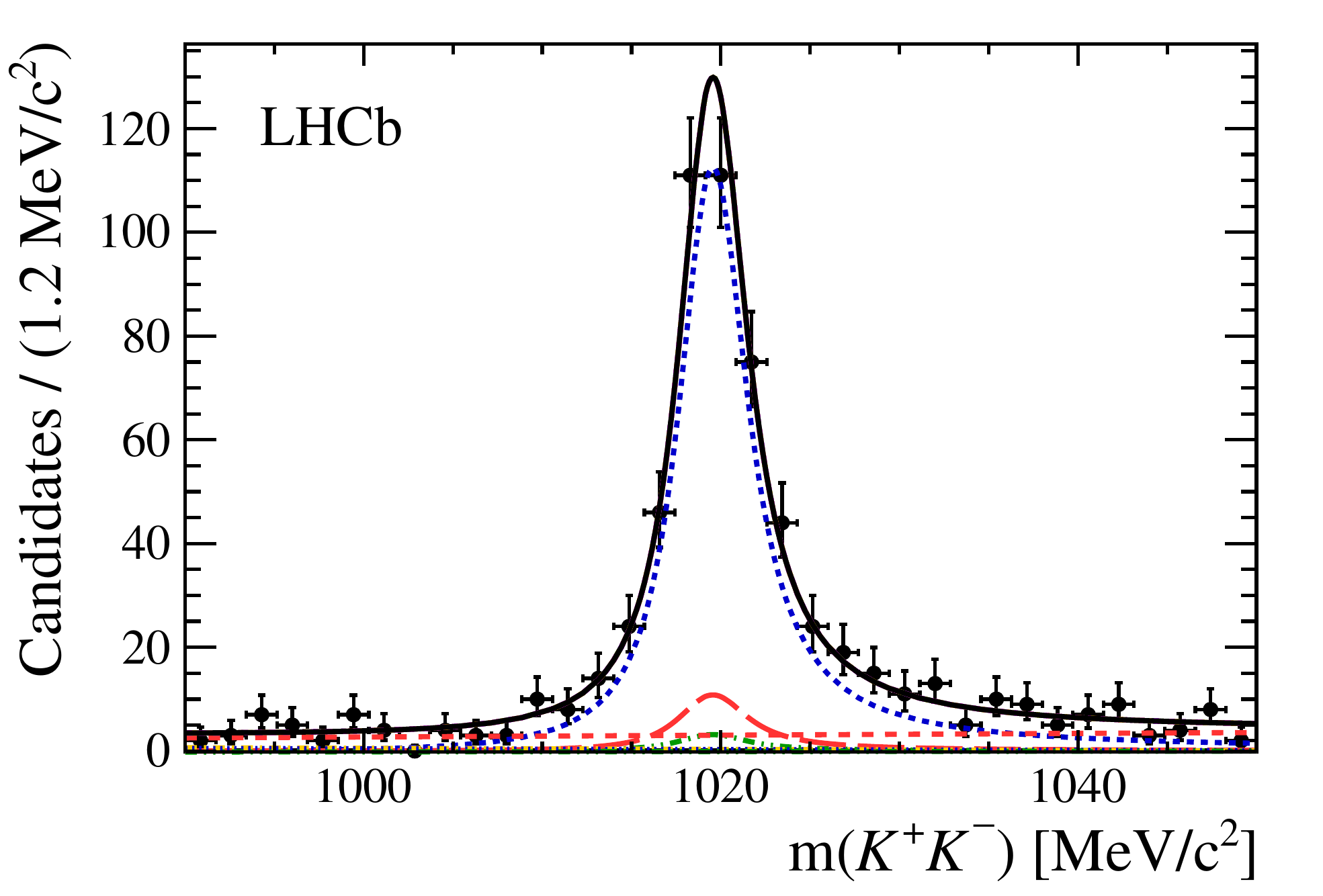}\\
\caption{\label{fig:fig6}
Distribution of the $p\bar{p}\Kp\Km$ (top) and $\Kp\Km$ (bottom) invariant masses. Points with error bars show the data.  
The solid black curve is the projection of the total fit result. 
The short-dashed and dotted blue curves show the $\Bs \to p\bar{p}\phi$ and $\Bs \to p\bar{p}\Kp\Km$ contributions, respectively. 
The long-dashed and medium-dashed red curves show the contributions of the combinatorial background with prompt $\phi$ and NR $\Kp\Km$, respectively. 
The dashed-double-dotted green, dashed-triple-dotted green and dashed-single-dotted-yellow curves show the $\Bd \to p\bar{p}\phi$, $\Bd \to p\bar{p}\Kp\Km$ and $\Bd \to p\bar{p}\Kp\pim$ contributions, respectively. 
}
\end{figure}

\begin{figure}[htbp]
\centering
\includegraphics[height=10.cm]{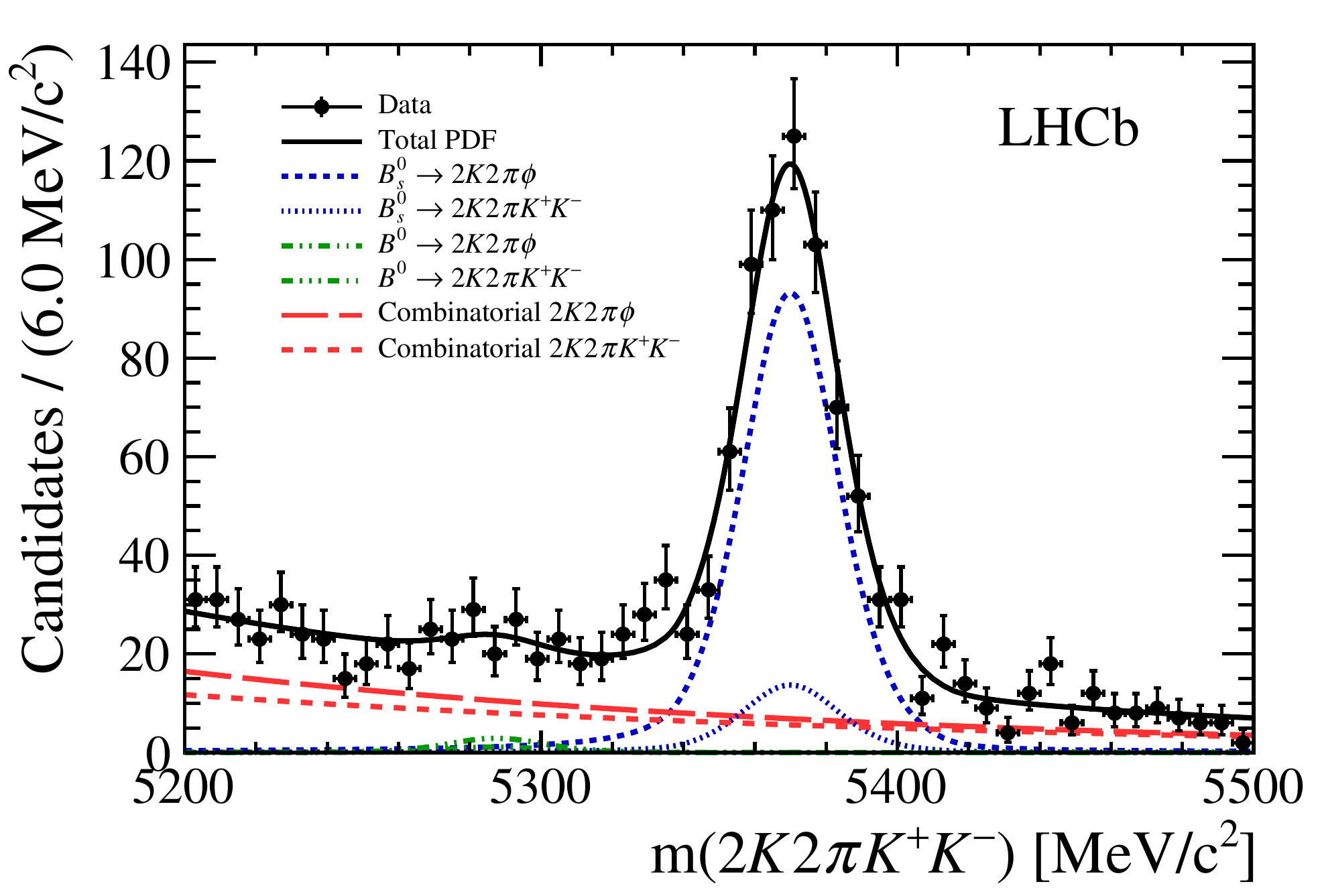}\\
\includegraphics[height=10.cm]{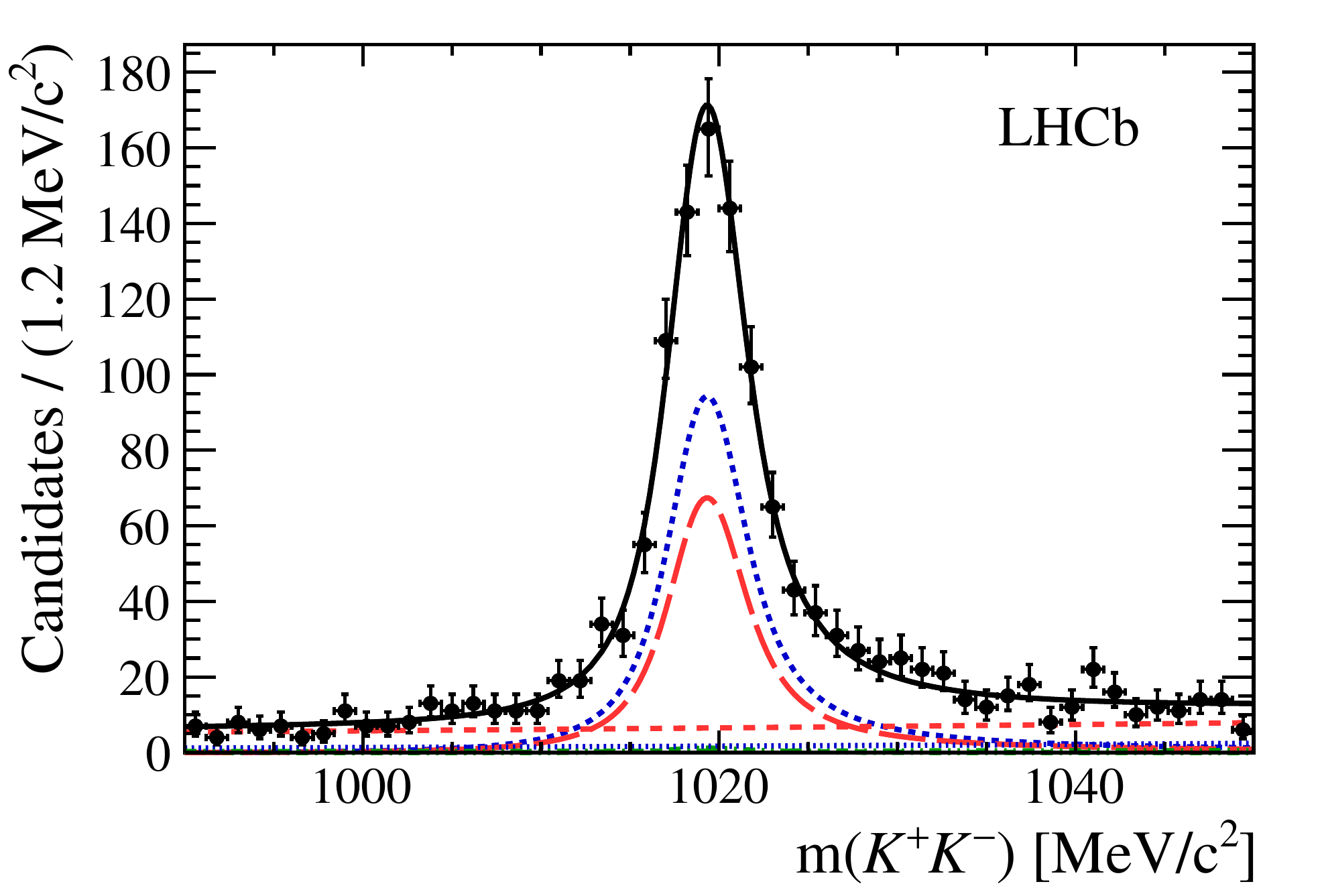}
\caption{\label{fig:2D_fit_2K2pi}
Distribution of the $2K2\pi\Kp\Km$ (top) and $\Kp\Km$ (bottom) invariant masses. Points with error bars show the data.  
The solid black curve is the projection of the total fit result. 
The short-dashed and dotted blue curves show the $\Bs \to 2K2\pi\phi$ and $\Bs \to 2K2\pi\Kp\Km$ contributions, respectively. 
The long-dashed and medium-dashed red curves show the contributions of the combinatorial background with prompt $\phi$ and NR $\Kp\Km$, respectively. 
The dashed-double-dotted and dashed-triple-dotted green curves show the $\Bd \to 2K2\pi\phi$ and $\Bd \to 2K2\pi\Kp\Km$ contributions, respectively.
}
\end{figure}

\begin{figure}[htbp]
\centering
\includegraphics[height=10.cm]{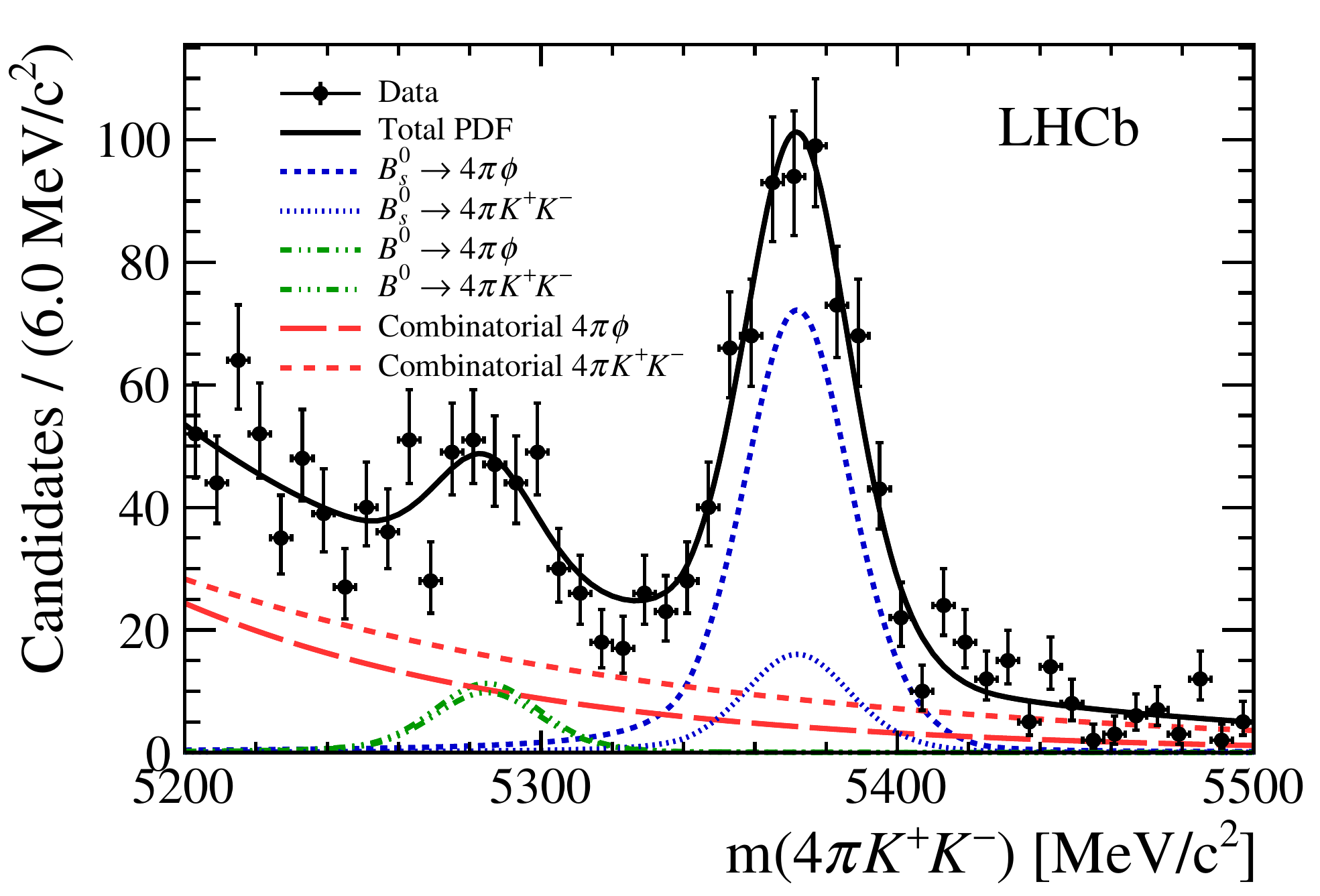}\\
\includegraphics[height=10.cm]{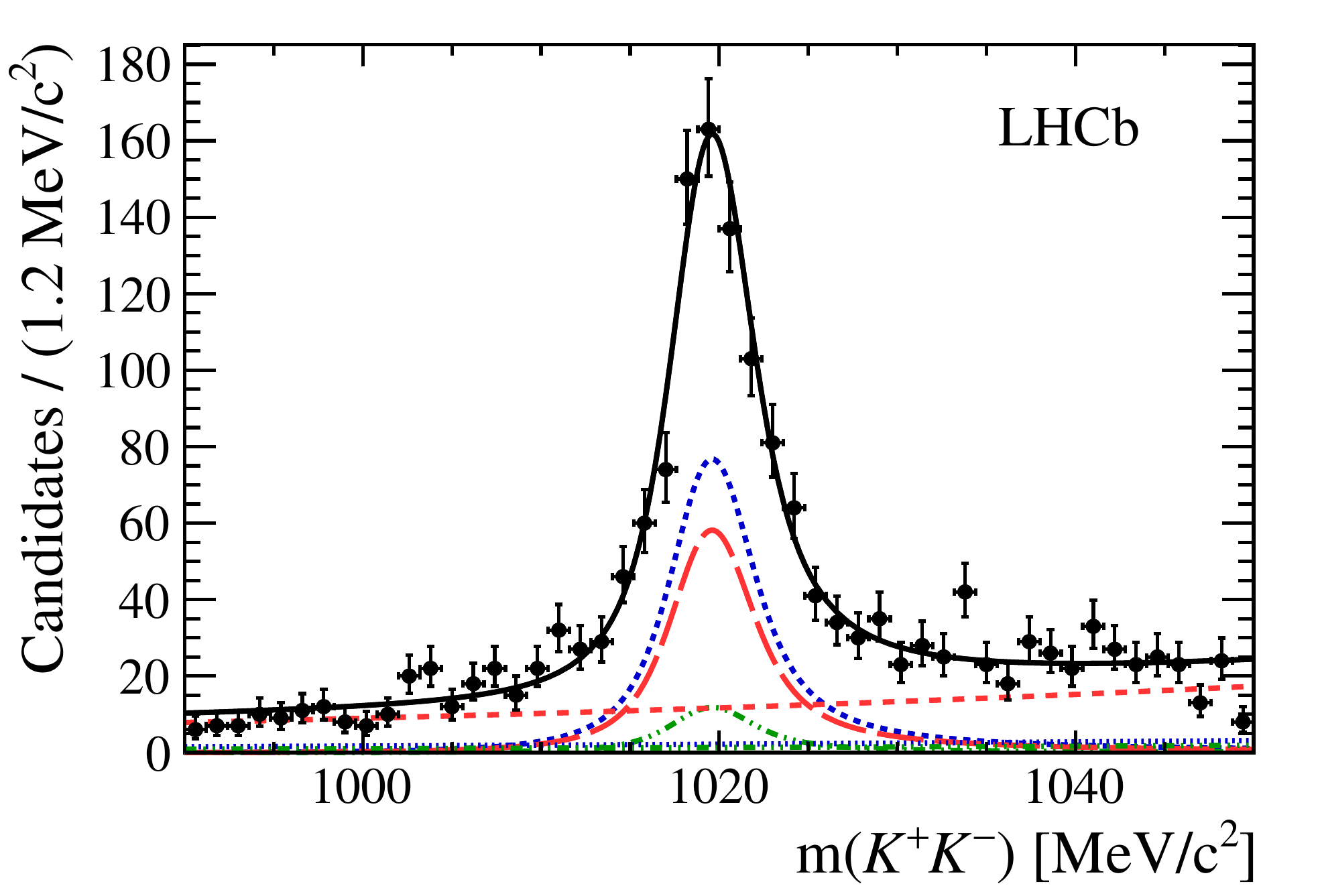}
\caption{\label{fig:2D_fit_4pi}
Distribution of the $4\pi\Kp\Km$ (top) and $\Kp\Km$ (bottom) invariant masses. Points with error bars show the data.  
The solid black curve is the projection of the total fit result. 
The short-dashed and dotted blue curves show the $\Bs \to 4\pi\phi$ and $\Bs \to 4\pi\Kp\Km$ contributions, respectively. 
The long-dashed and medium-dashed red curves show the contributions of the combinatorial background with prompt $\phi$ and NR $\Kp\Km$, respectively. 
The dashed-double-dotted and dashed-triple-dotted green curves show the $\Bd \to 4\pi\phi$ and $\Bd \to 4\pi\Kp\Km$ contributions, respectively.
}
\end{figure}

\begin{figure}[htbp]
\centering
\includegraphics[height=10.cm]{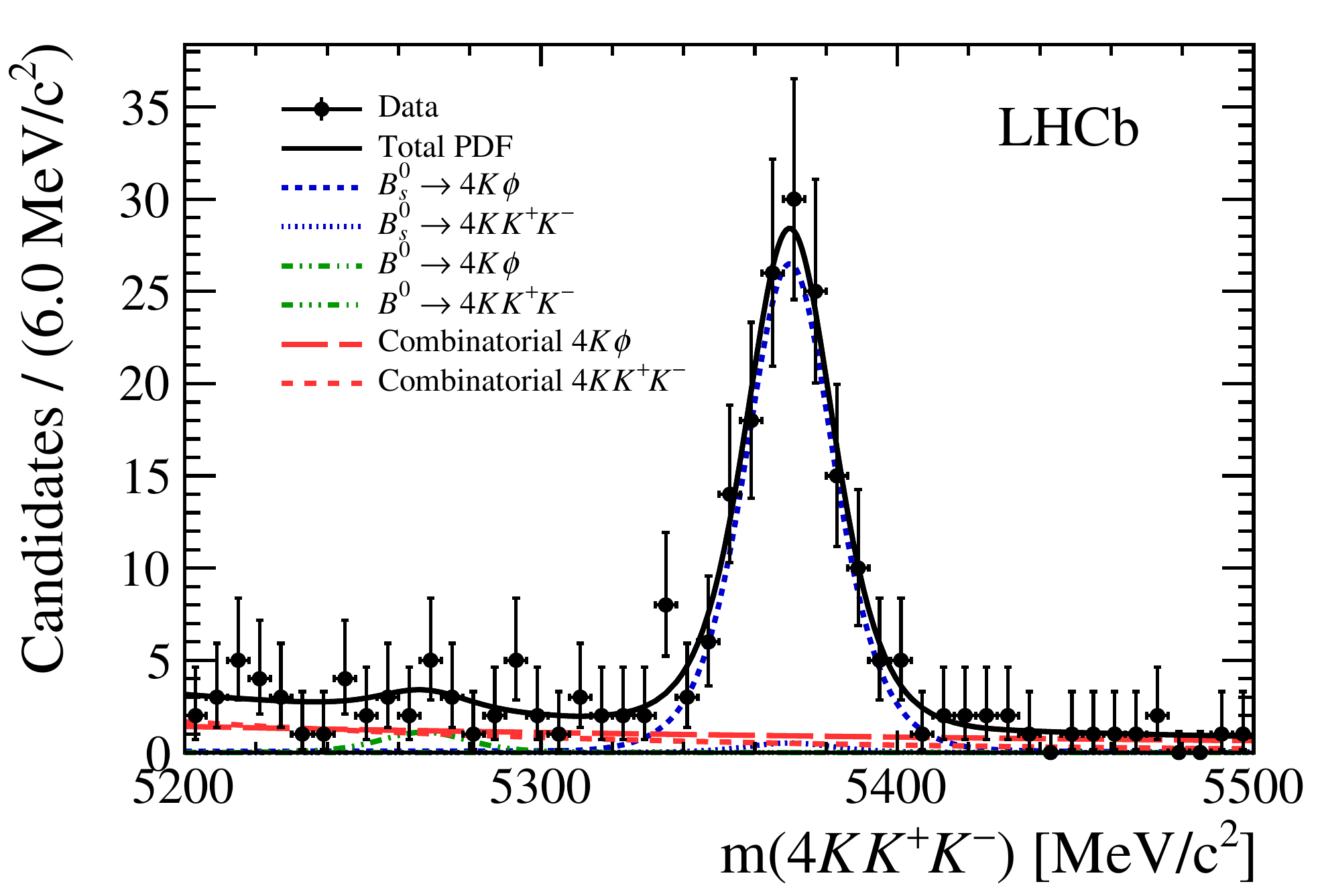}\\
\includegraphics[height=10.cm]{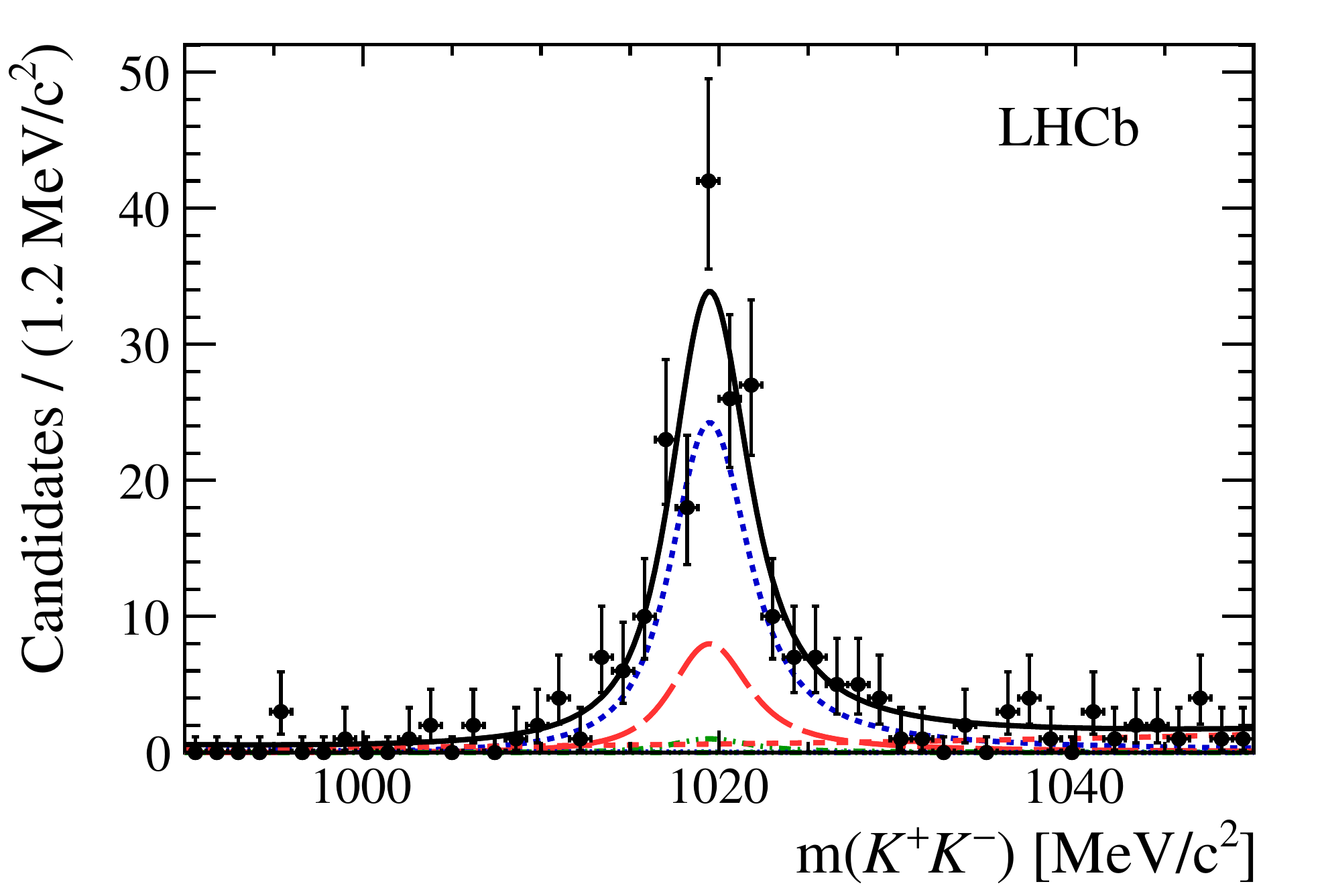}
\caption{\label{fig:fig9}
Distribution of the $4K\Kp\Km$ (top) and $\Kp\Km$ (bottom) invariant masses. Points with error bars show the data.  
The solid black curve is the projection of the total fit result. 
The short-dashed and dotted blue curves show the $\Bs \to 4K\phi$ and $\Bs \to 4K\Kp\Km$ contributions, respectively. 
The long-dashed and medium-dashed red curves show the contributions of the combinatorial background with prompt $\phi$ and NR $\Kp\Km$, respectively. 
The dashed-double-dotted and dashed-triple-dotted green curves show the $\Bd \to 4K\phi$ and $\Bd \to 4K\Kp\Km$ contributions, respectively.
}
\end{figure}

The $p\bar{p}\pip\pim$ invariant mass distribution and the fit projection are shown in Fig.~\ref{fig:fig5}.
The four $p\bar{p}(4h)\Kp\Km$ and $\Kp\Km$ invariant-mass distributions and the corresponding two-dimensional fit projections are shown in Figs.~\ref{fig:fig6} to~\ref{fig:fig9}.


\clearpage
\section{Correlation matrix}
\label{app:corrMatrix} 

The statistical correlation matrix for the simultaneous fit to the $p\bar{p}$ and $4h$ invariant-mass distributions for $\Bs \to p\bar{p}\phi$ and $\Bs \to 4h\phi$ candidates is given in Table~\ref{tab:amplitude_fit_simultaneous_corr}.

\begin{table}[h]
\caption{Statistical correlation matrix for the parameters from the simultaneous fit to the $p\bar{p}$ and $4h$ invariant-mass spectra for $\Bs \to p\bar{p}\phi$ and $\Bs \to 4h\phi$ candidates.}
\label{tab:amplitude_fit_simultaneous_corr}
\centering
\begin{tabular}{l | cc cc cc cc c } 
& $\kappa_{\rm NR}^{2K2\pi}$ & $\kappa_{\rm NR}^{4K}$ & $\kappa_{\rm NR}^{4\pi}$ & $\kappa_{\textrm{bkg}}^{2K2\pi}$ & $\kappa_{\textrm{bkg}}^{4K}$ & $\kappa_{\textrm{bkg}}^{4\pi}$ & $\xi_{\rm NR}^{2K2\pi}$ & $\xi_{\rm NR}^{4K}$ & $\xi_{\rm NR}^{4\pi}$  \\ 
 \hline 
$\mathcal{B}(B_{s}^{0} \to \etac \phi)$  &  $+0.22$ & $-0.00$ & $-0.05$ & $-0.19$ & $+0.00$ & $-0.00$ & $+0.55$ & $+0.04$ & $+0.13$  \\ 
 $\kappa_{\rm NR}^{2K2\pi}$       &  & $-0.00$ & $-0.01$ & $+0.65$ & $-0.00$ & $-0.00$ & $+0.07$ & $+0.01$ & $+0.02$  \\ 
 $\kappa_{\rm NR}^{4K}$           &  &  & $-0.01$ & $+0.00$ & $-0.06$ & $-0.00$ & $+0.00$ & $+0.37$ & $+0.02$  \\ 
 $\kappa_{\rm NR}^{4\pi}$         &  &  &  &  $-0.00$ & $-0.01$ & $+0.25$ & $-0.10$ & $+0.09$ & $-0.72$  \\ 
 $\kappa_{\textrm{bkg}}^{2K2\pi}$ &  &  &  &  &  $+0.00$ & $-0.00$ & $-0.47$ & $-0.01$ & $-0.01$   \\ 
 $\kappa_{\textrm{bkg}}^{4K}$     &  &  &  &  &  &  $-0.00$ & $+0.01$ & $-0.21$ & $+0.02$  \\ 
 $\kappa_{\textrm{bkg}}^{4\pi}$   &  &  &  &  &  &  &  $-0.01$ & $+0.01$ & $-0.07$  \\ 
 $\xi_{\rm NR}^{2K2\pi}$  &  &  &  &  &  &  &  &  $-0.02$ & $+0.17$  \\ 
 $\xi_{\rm NR}^{4K}$      &  &  &  &  &  &  &  &  &  $-0.12$   \\ 
 \cline{1-10} 
&\multicolumn{9}{c}{ }\\
& & $\xi_{\rm bkg}^{2K2\pi}$ & $\xi_{\rm bkg}^{4K}$ & $\xi_{\rm bkg}^{4\pi}$ & $m_{\etac}$ & $m_{\jpsi}$ & $\delta\varphi_{2K2\pi}$ & $\delta\varphi_{4K}$ & $\delta\varphi_{4\pi}$ \\
  \cline{1-10}
$\mathcal{B}(B_{s}^{0} \to \etac \phi)$ & & $-0.40$ & $-0.04$ & $-0.11$ & $-0.02$ & $+0.00$ & $-0.23$ & $+0.28$ & $-0.44$ \\
$\kappa_{\rm NR}^{2K2\pi}$ & & $-0.00$ & $-0.01$ & $-0.02$ & $+0.01$ & $+0.00$ & $-0.02$ & $+0.06$ & $-0.09$ \\
$\kappa_{\rm NR}^{4K}$     & & $-0.00$ & $-0.26$ & $-0.01$ & $-0.03$ & $-0.00$ & $+0.01$ & $-0.10$ & $+0.00$ \\
$\kappa_{\rm NR}^{4\pi}$  & & $+0.05$ & $-0.05$ & $+0.52$ & $+0.45$ & $-0.04$ & $-0.15$ & $+0.02$ & $+0.09$ \\
$\kappa_{\textrm{bkg}}^{2K2\pi}$ & & $+0.55$ & $+0.01$ & $+0.01$ & $-0.03$ & $+0.00$ & $+0.49$ & $-0.05$ & $+0.08$ \\
$\kappa_{\textrm{bkg}}^{4K}$ & & $-0.00$ & $+0.04$ & $-0.01$ & $-0.03$ & $-0.00$ & $+0.01$ & $-0.17$ & $-0.00$ \\
$\kappa_{\textrm{bkg}}^{4\pi}$ & & $+0.00$ & $-0.00$ & $-0.33$ & $+0.04$ & $-0.00$ & $-0.01$ & $+0.00$ & $+0.03$ \\
$\xi_{\rm NR}^{2K2\pi}$ & & $-0.82$ & $-0.00$ & $-0.12$ & $-0.18$ & $+0.01$ & $-0.23$ & $+0.14$ & $-0.24$ \\
$\xi_{\rm NR}^{4K}$ & & $-0.00$ & $-0.52$ & $+0.07$ & $+0.21$ & $-0.01$ & $-0.08$ & $+0.07$ & $-0.01$ \\
$\xi_{\rm NR}^{4\pi}$ & & $-0.09$ & $+0.06$ & $-0.49$ & $-0.61$ & $+0.01$ & $+0.18$ & $-0.01$ & $+0.05$ \\
$\xi_{\rm bkg}^{2K2\pi}$ & &  &  $+0.01$ & $+0.07$ & $+0.07$ & $-0.02$ & $+0.41$ & $-0.11$ & $+0.18$ \\
$\xi_{\rm bkg}^{4K}$ & &  &  &  $-0.03$ & $-0.11$ & $-0.02$ & $+0.05$ & $-0.26$ & $+0.01$ \\
$\xi_{\rm bkg}^{4\pi}$ & &  &  &  &  $+0.34$ & $-0.02$ & $-0.09$ & $-0.01$ & $+0.28$  \\
$m_{\etac}$  & &  &  &  &  &  $-0.01$ & $-0.35$ & $+0.07$ & $+0.02$ \\
$m_{\jpsi}$ & &  &  &  &  &  &  $+0.01$ & $+0.01$ & $-0.01$ \\
$\delta\varphi_{2K2\pi}$ & &  &  &  &  &  &  &  $-0.09$ & $+0.10$ \\
$\delta\varphi_{4K}$ & &  &  &  &  &  &  &  &  $-0.12$ \\
  \cline{1-10} 

\end{tabular}
\end{table}

\cleardoublepage
\newpage
\addcontentsline{toc}{section}{References}
\setboolean{inbibliography}{true}
\bibliographystyle{LHCb}
\bibliography{main,LHCb-PAPER,LHCb-CONF,LHCb-DP,LHCb-TDR}

\newpage
\centerline{\large\bf LHCb collaboration}
\begin{flushleft}
\small
R.~Aaij$^{40}$,
B.~Adeva$^{39}$,
M.~Adinolfi$^{48}$,
Z.~Ajaltouni$^{5}$,
S.~Akar$^{59}$,
J.~Albrecht$^{10}$,
F.~Alessio$^{40}$,
M.~Alexander$^{53}$,
S.~Ali$^{43}$,
G.~Alkhazov$^{31}$
% ...

{\footnotesize \it
% ...
$ ^{5}$Clermont Universit{\'e}, Universit{\'e} Blaise Pascal, CNRS/IN2P3, LPC, Clermont-Ferrand, France\\
% ...
$ ^{10}$Fakult{\"a}t Physik, Technische Universit{\"a}t Dortmund, Dortmund, Germany\\
% ...
$ ^{31}$Petersburg Nuclear Physics Institute (PNPI), Gatchina, Russia\\
% ...
$ ^{39}$Universidad de Santiago de Compostela, Santiago de Compostela, Spain\\
$ ^{40}$European Organization for Nuclear Research (CERN), Geneva, Switzerland\\
% ...
$ ^{43}$Nikhef National Institute for Subatomic Physics, Amsterdam, The Netherlands\\
% ...
$ ^{48}$H.H. Wills Physics Laboratory, University of Bristol, Bristol, United Kingdom\\
% ...
$ ^{53}$School of Physics and Astronomy, University of Glasgow, Glasgow, United Kingdom\\
% ...
$ ^{59}$University of Cincinnati, Cincinnati, OH, United States\\
% ...
}
\end{flushleft}

\end{document}